\documentclass[11pt]{emulateapj}
\usepackage{graphicx}
\usepackage{amssymb,amsmath}
\usepackage{natbib}
\begin{document}
\topmargin 0.5in 

\begin{abstract}
We study the relationship between the field star formation and cluster formation properties in a large sample of nearby dwarf galaxies. We use optical data from the Hubble Space Telescope and from ground-based telescopes to derive the ages and masses of the young ($t_{\rm{age}} \lesssim 100~$Myr) cluster sample. Our data provides the first constraints on two proposed relationships between the star formation rate of galaxies and the properties of their cluster systems in the low star formation rate regime. The data show broad agreement with these relationships, but significant galaxy-to-galaxy scatter exists.  In part, this scatter can be accounted for by simulating the small number of clusters detected from stochastically sampling the cluster mass function.  However, this stochasticity does not fully account for the observed scatter in our data suggesting there may be true variations in the fraction of stars formed in clusters in dwarf galaxies. Comparison of the cluster formation and the brightest cluster in our sample galaxies also provide constraints on cluster destruction models.

\end{abstract}

\title{The ACS Nearby Galaxy Survey Treasury. X. Quantifying the Star Cluster Formation Efficiency of Nearby Dwarf Galaxies}
\author{David O. Cook\altaffilmark{1}}
\author{Anil C. Seth\altaffilmark{5}}
\author{Daniel A. Dale\altaffilmark{1}}
\author{L. Clifton Johnson\altaffilmark{3}}
\author{Daniel R. Weisz\altaffilmark{3}}
\author{Morgan Fouesneau\altaffilmark{3}}
\author{Knut A. G. Olsen\altaffilmark{4}}
\author{Charles W. Engelbracht\altaffilmark{2}}
\author{Julianne J. Dalcanton\altaffilmark{3}}
\altaffiltext{1}{Department of Physics \& Astronomy, University of Wyoming, Laramie, WY 82071, USA; dcook12$@$uwyo.edu}
\altaffiltext{2}{Steward Observatory, University of Arizona, Tucson, AZ 85721, USA}
\altaffiltext{3}{Department of Astronomy, University of Washington, Seattle, WA 98195, USA}
\altaffiltext{4}{National Optical Astronomy Observatory, Tucson, AZ 85719, USA}
\altaffiltext{5}{Department of Physics \& Astronomy, University of Utah, Salt Lake City, UT 84112, USA}
\maketitle

\section{Introduction}
In galaxies, the relationship between stars in bound clusters and those that populate the field is poorly understood. The formation of both populations is connected, whether stars form in clusters and dissolve to create field stars, or field stars form concurrently with clusters. Until recently, it was widely accepted that the majority of stars form in a cluster environment \citep{lada03,porras03}. However, infrared surveys have shown evidence that star formation occurs in a continuum of stellar densities, including formation of unbound field stars \citep[e.g.,][]{evans09,Bressert10}.  We study the behavior of field and clustered star formation in dwarf galaxies by quantifying the fraction of stars that populate star clusters.  

A quantity that directly probes the fraction of stars located in clusters is the Cluster Formation Efficiency ($\Gamma$), defined as the Cluster Formation Rate divided by the Star Formation Rate \citep[CFR/SFR;][]{Bastian08}. This quantity has been observed to positively correlate with the star formation rate density $\Sigma_{SFR}$ (SFR/kpc$^2$), suggesting that higher star formation rates provide an environment beneficial for cluster formation. \cite{goddard10} first found this trend to be a power-law relationship for spirals and starburst dwarf galaxies \citep[see also][]{larsenrichtler00}. Later, \cite{adamo11} found that the correlation extended to galaxies with extreme star formation (e.g., starburst nuclei and blue compact galaxies). However, both \cite{Bastian08} and \cite{villaLarsen11} found no clear trend in $\Gamma$ with the star formation rate in their studies of spiral and starbursting galaxies. In this paper we will test whether this trend is seen at lower SFR densities and whether significant scatter exists around the trend. 

Determination of the star cluster formation efficiency in low SFR density galaxies has complications due to cluster populations which are dominated by low mass clusters. Recent studies have found that estimated cluster parameters when derived by comparing observed photometry to synthetic spectra show a large spread in derived ages and masses for clusters below a few times 10$^{4}$ M$_{\odot}$ \citep[e.g.,][]{anders04a,piskunov09,fouesneau10,bogdan10b,villaLarsen11}. The spread in derived ages and masses results from the random presence or absence of a few post main-sequence stars that dominate the light of the cluster. For instance, a young cluster of 10$^4$ M$_{\odot}$ is expected to have on average one supergiant, but could end up with multiple or no supergiants. As a result, adding a luminous red supergiant to the average prediction will shift the fitted population towards older ages. Bayesian methods are now available that can explicitly take into account these stochastic fluctuations of integrated colors and fluxes of star clusters \citep{fouesneau10}. These effects need to be taken into account when determining the cluster formation rate of low mass systems.

Another complication of measuring $\Gamma$ for our sample is due to the inconsistency of different star formation rate indicators in dwarf galaxies. A galaxy's SFR can be estimated with many different tracers (H$\alpha$, integrated far-infrared, UV, resolved star Color-Magnitude Diagram (CMD) fitting, etc.).  Unfortunately, the integrated flux tracers, although useful, yield SFR approximations based on stars of many ages and become inconsistent with each other in low mass galaxies \citep{meurer09,lee09b,boselli09,hunter10,weisz12}. Ideally, knowledge of every individual star's age and mass would provide an accurate SFR. However, the closest method to this is reconstructing star formation histories using individually resolved stars in a galaxy. Utilizing HST's high image resolution capabilities, the star formation histories can be calculated in nearby galaxies ($D < 10 $Mpc). Impressive temporal resolution at young ages ($t_{\rm{age}}< 100~$Myr) can be obtained and can provide an excellent foundation against which young cluster ages can be compared \citep[][]{weisz08,mcquinn11,villaLarsen11}. The mass and age distributions of star clusters combined with CMD-derived field star formation histories can produce a direct measurement of the number of stars that remain in bound star clusters, $\Gamma$.  

Despite the the difficulties associated with analyzing dwarf galaxy properties, there are many advantages. Dwarf galaxies are abundant in the local universe and lack spiral structure. Also, low extinction plus low surface brightness make finding star clusters relatively easy. While few clusters are formed in individual low star formation rate dwarfs, the large number of local dwarf galaxies enables construction of a statistically significant cluster sample from many galaxies. These properties make dwarfs an ideal laboratory for identifying star clusters and studying individual resolved stars.  

In this work we will test existing relationships found between the cluster and field star populations at young ($t_{\rm{age}} \lesssim 100~$Myr) ages for a sample of low star formation rate density, dwarf galaxies, and we examine in detail if the cluster sample is consistent with a universal cluster formation mechanism.
  
\section{Data}
In this section we present the data and analysis used to derive cluster and star formation rates in a large sample of dwarf galaxies.  We identify clusters using HST data, and derive their ages and masses using a combination of HST and ground-based photometry.  We also use the HST imaging of individual stars to derive recent star formation histories of each galaxy.

\subsection{Sample \& Observations}
The sample consists of 37 dwarf galaxies selected from the ACS Nearby Galaxy Survey Treasury (ANGST) project of \cite{dalcanton09}. ANGST is a volume-limited survey and and therefore it contains a large number of dwarf galaxies.  We restricted the ANGST sample to galaxies fainter than $M_{B}=-18$ in order eliminate non-dwarf galaxies.  We further restricted the sample to contain galaxies that have available Advanced Camera for Surveys (ACS) data (some ANGST data was taken with WFPC2), and that are observable from the northern hemisphere.  All of these restrictions produce a sample of 37 dwarf galaxies from the original 69 of the ANGST parent sample; properties of the final sample can be found in Table 1. The distance range of the sample is $\sim$2--4.5 Mpc; at these distances the majority of star clusters are partially resolved into individual stars at the resolution of HST.  Crowding issues near the core of the clusters result in an incomplete census of the resolved stars and therefore make it impossible to derive accurate cluster ages and masses from CMD fitting methods.  Therefore, we supplemented ANGST data with ground-based data to obtain a Johnson $UBV$ and Cousins $RI$ data set for each galaxy in our sample that contained clusters, thereby enabling accurate integrated-light age and mass determinations.  

The majority of galaxies in our sample have ANGST ACS Wide Field Camera (WFC) data taken in the F555W/F606W ($\sim$$V$) and F814W ($\sim$$I$).  We therefore obtained ground based data in the $U$, $B$ and $R$ filters.  The $UBR$ ground-based data came from both the BOK telescope atop Kitt Peak and WIRO. The 2.1 meter BOK telescope, operated by Steward Observatory, in the 90 prime configuration \citep{williams04} has a blue optimized Anti-Reflective (AR) coated CCD which facilitates deep $U$ images with significantly shorter exposure times than with traditional non-coated CCDs. The BOK data were taken over three runs: November 2009, May 2010, and December 2010. The remaining $B$ and all of the $R$ data were taken at the WIRO 2.3 meter telescope in the prime focus configuration over the months of April 2010, March 2010, September 2010, March 2011, and July 2011. Two galaxies, DDO78 and UGCA292, have ACS WFC data in the F475W and F814W  filters.  For these galaxies, we obtained broadband $V$ images of these galaxies at the Wyoming InfraRed Observatory (WIRO) in order to complete the $UBVRI$ data set. There are also two galaxies whose full data set was not obtained (GR8 and UGC5442); these galaxies are not included in the analysis.

\begin{center}
\begin{deluxetable*}{lclrrrrccc}[Hb]
\tabletypesize{\footnotesize}
\tablecolumns{9} 
\tablewidth{0pt}
\tablecaption{Dwarf Galaxy Parameters}
\tablehead{\colhead{Galaxy}        	& 
           \colhead{Alternative}	& 
           \colhead{RA}       		& 
           \colhead{Dec}      		&
           \colhead{$a$}      		&
           \colhead{$b$}      		&
           \colhead{$M_B$}              &
           \colhead{Dist}         	&
           \colhead{12+log(O$/$H)}      \\
           				&
           \colhead{Name}		& 
           \colhead{(J2000.0)} 		&        
           \colhead{(J2000.0)}		&
           \colhead{($^{\prime\prime}$)}		&
           \colhead{($^{\prime\prime}$)}		&
           \colhead{(mag)}		&
	   \colhead{(Mpc)}		&
                                        \\
           \colhead{(1)}		& 
           \colhead{(2)}		& 
           \colhead{(3)} 		&        
           \colhead{(4)}		&
           \colhead{(5)}		&
           \colhead{(6)}		&
           \colhead{(7)}		&
           \colhead{(8)}		& 
	   \colhead{(9)}		} 

\startdata
ESO540-G030   & KDG2 		& 00 49 21.1 & $-$18 04 28 & 84  & 74  & $-$11.29 & 3.33 & \\  
NGC404	      & UGC0718 	& 01 09 26.9 &  35 43 03 & 105 & 105 & $-$16.25 & 3.05 & \\  
KKH37	      &  		& 06 47 45.8 &  80 07 26 & 52  & 40  & $-$11.26 & 3.26 & \\  
NGC2366	      & UGC3851,DDO42 	& 07 28 52.0 &  69 12 19 & 308 & 158 & $-$15.85 & 3.21 & 7.95 $\pm$ 0.05\tablenotemark{b} \\  
UGCA133	      & DDO44 		& 07 34 11.3 &  66 53 10 & 108 & 76  & $-$11.89 & 3.10 & \\  %
UGC4305	      & HoII,DDO50 	& 08 19 05.9 &  70 42 51 & 278 & 232 & $-$16.57 & 3.38 & 7.72 $\pm$ 0.14\tablenotemark{a} 7.92 $\pm$ 0.10\tablenotemark{b} \\ 
M81dwA	      & KDG52           & 08 23 56.0 &  71 01 46 & 39  & 39  & $-$11.37 & 3.44 & \\  
UGC4459	      & DDO53           & 08 34 06.5 &  66 10 45 & 67  & 56  & $-$13.23 & 3.61 & 7.60 $\pm$ 0.11\tablenotemark{a} 7.82 $\pm$ 0.09\tablenotemark{b} \\  
UGC5139	      & HoI,DDO63 	& 09 40 28.2 &  71 11 11 & 132 & 110 & $-$14.26 & 3.90 & 7.60 $\pm$ 0.11\tablenotemark{a} 8.00 $\pm$ 0.10\tablenotemark{b} \\  
$[$FM2000$]$ 1     &  		& 09 45 25.6 &  68 45 27 & 44  & 44  & $-$10.16 & 3.53 & \\  
BK3N	      &  		& 09 53 48.5 &  68 58 09 & 20  & 20  & $-$9.23  & 3.86 & \\  
KDG61	      &  		& 09 57 02.7 &  68 35 30 & 107 & 60  & $-$12.54 & 3.49 & 8.35 $\pm$ 0.05\tablenotemark{b} \\  
Arp's loop    &  		& 09 57 29.0 &  69 16 20 & 68  & 68  & $-$11.16 & 3.78 & \\  
UGC5336	      &HoIX,DDO66,KDG62 & 09 57 32.4 &  69 02 35 & 124 & 90  & $-$13.31 & 3.61 & 8.14 $\pm$ 0.11\tablenotemark{a} 8.65 $\pm$ 0.25\tablenotemark{b} \\  
LEDA166101    & KK98\_77        & 09 50 10.0 &  67 30 24 & 110 & 76  & $-$11.42 & 3.55 & \\  
UGC5428	      & DDO71,KDG63     & 10 05 07.3 &  66 33 18 & 98  & 84  & $-$11.71 & 3.53 & 7.4\tablenotemark{c}\\  
UGC5442	      & KDG64 	        & 10 07 01.9 &  67 49 39 & 98  & 62  & $-$12.32 & 3.72 & 7.4\tablenotemark{c}\\  
IKN	      &  		& 10 08 05.9 &  68 23 57 & 90  & 78  & $-$10.84 & 3.61 & 7.3\tablenotemark{c}\\  
$[$HS98$]$ 117     &  		& 10 21 25.2 &  71 06 58 & 106 & 64  & $-$11.51 & 3.82 & \\  
DDO78	      &  		& 10 26 27.9 &  67 39 24 & 70  & 70  & $-$12.04 & 3.66 & 7.4\tablenotemark{c}\\  
IC2574	      & UGC5666,DDO81   & 10 28 22.4 &  68 24 58 & 432 & 243 & $-$17.17 & 3.81 & 7.85 $\pm$ 0.14\tablenotemark{a} \\  
UGC5692	      & DDO82 	        & 10 30 35.0 &  70 37 10 & 153 & 106 & $-$14.44 & 3.80 & 7.95 $\pm$ 0.20\tablenotemark{b} \\  
KDG73	      &  	        & 10 52 55.3 &  69 32 45 & 63  & 50  & $-$10.75 & 4.03 & \\  
NGC3741	      & UGC6572         & 11 36 06.4 &  45 17 07 & 100 & 54  & $-$13.01 & 3.24 & 7.62 $\pm$ 0.20\tablenotemark{b} \\  
NGC4163	      & NGC4167,UGC7199 & 12 12 08.9 &  36 10 10 & 106 & 82  & $-$13.76 & 2.87 & 7.91 $\pm$ 0.20\tablenotemark{b} \\  
UGCA276	      & DDO113,KDG90    & 12 14 57.9 &  36 13 08 & 80  & 72  & $-$11.61 & 2.95 & \\  
UGCA292	      &  		& 12 38 40.0 &  32 46 00 & 58  & 38  & $-$11.36 & 3.62 & 7.30 $\pm$ 0.03\tablenotemark{b} \\  
UGC8091	      & GR8  		& 12 58 40.4 &  14 13 03 & 62  & 46  & $-$12.00 & 2.08 & 7.65 $\pm$ 0.06\tablenotemark{b} \\    
UGC8201	      & DDO165          & 13 06 26.8 &  67 42 15 & 134 & 75  & $-$15.09 & 4.57 & 7.63 $\pm$ 0.08\tablenotemark{a} 7.80 $\pm$  0.20\tablenotemark{b} \\  
UGC8508	      &                 & 13 30 44.4 &  54 54 36 & 80  & 60  & $-$12.95 & 2.58 & 7.89 $\pm$ 0.20\tablenotemark{b} \\  
UGC8651	      & DDO181          & 13 39 53.8 &  40 44 21 & 97  & 70  & $-$12.94 & 3.14 & 7.85 $\pm$ 0.04\tablenotemark{b} \\  
UGC8760	      & DDO183          & 13 50 51.1 &  38 01 16 & 108 & 56  & $-$13.08 & 3.22 & 7.6\tablenotemark{c}\\  
UGC8833	      &                 & 13 54 48.7 &  35 50 15 & 60  & 58  & $-$12.31 & 3.08 & \\  
KKR03	      &                 & 14 07 10.7 &  35 03 37 & 34  & 24  & $-$8.49  & 1.97 & \\  
UGC9128	      & DDO187          & 14 15 56.5 &  23 03 19 & 64  & 44  & $-$12.43 & 2.21 & 7.75 $\pm$ 0.05\tablenotemark{b} \\  
UGC9240	      & DDO190          & 14 24 43.5 &  44 31 33 & 111 & 91  & $-$14.14 & 2.79 & 7.95 $\pm$ 0.03\tablenotemark{b} \\  
KKH98	      &  		& 23 45 34.0 &  38 43 04 & 63  & 40  & $-$10.29 & 2.54 & \\  
\enddata
\tablecomments{Properties of the Sample Dwarf galaxies--Columns 1 and 2: galaxy names;Column 3 and 4: J2000 Right Ascension (RA) and Declination (DEC) as reported by the NASA/IPAC Extragalactic Database (NED);Columns 5 and 6: semi-major and semi-minor axis measured by the 3.6 $\mu$m images by \cite{dale09};Column 7: absolute B magnitude \citep{karachentsev04};Column 8: adopted distance (Tip of the Red Giant Branch method) \citep{dalcanton09};Column 9: metallicity taken from different sources: $^a$\cite{moustakas10}, $^b$\cite{marble10}, $^c$\cite{hlee07}.}
\end{deluxetable*}
\end{center}
\vspace{-1cm}

\subsection{Cluster Identification}
The cluster catalog for the full ANGST sample will be presented in a separate work (Johnson et al. 2012; submitted); we briefly summarize how clusters were identified here. Cluster identification was performed by four members of our team via visual inspection of HST color images. Potential objects were chosen based on grouping of stars, spherical symmetry, colors that are consistent with single stellar population models \citep{marigo08,girardi10}, and a lack of structure (spiral arms or otherwise) that might indicate a background galaxy. However, since a cut based on colors that are inconsistent with non-stochastic stellar models could potentially remove young, low-mass clusters from the sample, we chose only to cut clusters who's color severely deviate (greater than one magnitude) from the model colors. After initial candidates were chosen from the images, all candidates were re-examined and ranked according to the likelihood that they were bound star clusters. Background elliptical galaxies are the most likely contaminant in our sample as these can be difficult to distinguish from older clusters if they fall on the face of the dwarf galaxy.  However, since we only aim to constrain the formation properties of young clusters ($t_{\rm{age}} \lesssim 100~$Myr), the red colors of potential background galaxies will typically exclude these objects from the final sample of young clusters. 

In total, we detect 144 clusters, 44 of which have an age less than 100~Myr in our 37 galaxy sample. Conversely, 20 galaxies in our sample contain no clusters at any age. We have tabulated the photometric properties of the clusters with ages less than 100~Myr in Table 4 located in the appendix. We include this table to facilitate comparison of our cluster fitting methods (discussed in Section 2.4) to others' methods.

\subsection{Photometry}
To use the ground-based and HST images as a uniform data set the HST images needed to be smoothed and re-binned to the ground-based images' resolution.  This step is necessary to achieve consistent magnitudes and colors because the HST images contain less light from nearby contaminating sources. Note that all of the clusters are unresolved in our ground-based images.

The smoothing was performed using the IDL routine FILTER\_IMAGE which convolves the image using a Gaussian kernel.  The FWHM of the Gaussian kernel was set to the typical seeing of the ground-based images of 1$\farcs$5. The re-binning was carried out via the IDL routine HASTROM and has the effect of enlarging the pixel size of space-based images to facilitate a comparison with the ground-based data.    

The cluster photometry was carried out in IDL with the routine APER, which was adapted from DAOPHOT.  Each cluster's photometry parameters were set via visual inspection of radial plots from the IRAF task PHOT in the interactive mode.  Clusters in non-crowded regions were given an aperture of twice the PSF FWHM and clusters in crowded regions were given aperture radii just below the radius at which contaminating light becomes prevalent.  To measure the local background, annuli were set at least 2 pixels from the edge of the photometric aperture. The pixel scale of the ground-based images and the convolved HST images is 0.5 arcseconds per pixel. 

Aperture photometry was performed on at least 10 unsaturated field stars at the two apertures and then at 5 times the FWHM.  The average and standard deviation of the flux ratios were used as the aperture correction and its uncertainty.  

HST zeropoints were obtained from the Space Telescope Science Institute website.\footnote{http://www.stsci.edu/hst/acs/analysis/zeropoints} After the recovery of the ACS camera on July 4, 2006 the operating temperature changed and resulted in a revised set of photometric zeropoints.  The appropriate zeropoints used were based on the observation date of each image. The HST data were then transformed into the Johnson-Cousins magnitude system via the prescription of \cite{sirianni05}.  We corrected for Galactic extinction using the \cite{schlegel98} maps tabulated by the NASA/IPAC Extragalactic Database.\footnote{http://ned.ipac.caltech.edu/}

\begin{figure}[h]
  \begin{center}
  \includegraphics[scale=0.6]{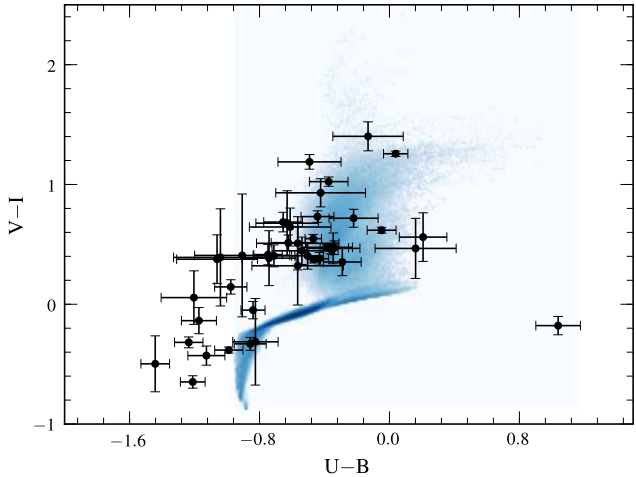}
  \caption{The color-color distribution of the cluster sample with ages less than 100~Myr . The shaded areas indicate the density of the stochastic models covering ages less than 100~Myr with Z=0.004 (SMC). The models lie primarily on a locus, but with a plume to red colors where red super giants have evolved off the main sequence.}
  \label{fig:figure1}
   \end{center}
\end{figure}  

\subsection{Cluster Properties}
Cluster ages and masses were determined using the Bayesian method of \cite{fouesneau10}. The method is based on a large collection of Monte-Carlo simulations of individual clusters, each containing a finite number of stars, which explicitly accounts for the stochastic effects induced by the random presence of luminous stars.  The synthetic clusters are constructed with the population synthesis code P\'EGASE.2 \citep{fioc99}. The underlying stellar evolution tracks are those of the Padova group \citep{bressan93}, with a simple extension through thermally pulsating Asymptotic Giant Branch (AGB) based on the prescriptions of \cite{groenewegen93}. The input stellar spectra are taken from the library of \cite{lejeune97}. The stellar Initial Mass Function (IMF) is taken from \cite{kroupa93}, and extends from 0.1 to 120 M$_{\odot}$. Nebular emission (lines and continuum) is included in the spectra and broadband fluxes under the assumption that no ionizing photons escape. 

The method is explained in detail by \cite{fouesneau10}, we therefore only briefly summarize the process here. Discrete model clusters will have a range of luminosities for a given mass due to the stochastic population of luminous stars. Hence mass-to-light ratios can not be treated as a constant for a cluster of a given age. For this reason, the total mass of a cluster cannot be determined by a simple scaling factor of the total light, and must be simultaneously fit along with other cluster parameters. The Bayesian method establishes joint probability distributions of the intrinsic cluster parameters (age, mass, metallicity, extinction), given a set of photometric measurements and their uncertainties. Each intrinsic property of a cluster is estimated from the comparison of its whole set of absolute fluxes to the distribution of a model catalog. The set of intrinsic properties with the largest probability is then assigned to the cluster.  




Although this method is more advanced than a $\chi^2$ minimization, it still incorporates multiple free parameters (e.g., age, mass, metallicity, internal extinction, etc). Independently quantifying some of these parameters or using reasonable assumptions can greatly increase the accuracy of derived cluster properties. We make two simplifying assumptions: (1) that the extinction is negligible in these low mass dwarf galaxies and (2) we fix the metallicity based on estimates of the current gas phase metallicity. 

The lack of internal extinction in dwarfs has been demonstrated by many local volume surveys, which find lower infrared to ultraviolet ratios for a given ultraviolet color \citep[e.g.,][]{buat05,dale07,munoz09,dale09}.  Additionally, the well defined main sequence and blue helium burning populations seen in the CMDs of these galaxies show that the internal reddening is small even for young stellar populations \citep{weisz08,weisz11b}. 

Fortunately, the metallicity of many of our galaxies have been observationally determined via spectroscopic studies of H$\rm II$ regions; listed in Table 1 \citep{marble10,moustakas10}.  The current available models of \cite{fouesneau10} have SMC, LMC, and Solar values (log(O/H)+12$\sim$8.0, 8.3, and 8.69, respectively) available at this time.  The majority of our galaxies with clusters show similar metallicities to that of the SMC, however two galaxies (KDG61 and UGC5336) have values similar to the LMC.  There are five galaxies (DDO 78, IKN, UGC5442, UGC5428, \& UGC8760) with clusters that do not have reliable metallicity measurements: these are fit with an SMC model. Using the Luminosity-Metallicity relationship of nearby galaxies ($z <$ 0.1) we have estimated that all five of these galaxies have a metallicity of 12+log(O/H)$\sim$7.5, given the absolute B magnitude of the galaxy \citep{hlee03,hlee07}. The mass in clusters of these galaxies will be biased to somewhat higher masses by approximately 5\% given the Padova SSP models.

The age ranges we choose to study are 4--10~Myr ($<$10~Myr) and 4--100~Myr ($<$100~Myr). We selected these age ranges to facilitate a direct comparison between our data and those of previous studies of cluster-host relationships. Furthermore, these age ranges are characteristic of the time scales of the $H\alpha$ and FUV SFR indicators ($<$10 and $<$100~Myr, respectively). The shorter age range ($<$10~Myr) mimics the age ranges of \cite{goddard10} and \citep{adamo11} while the longer age range ($<$100~Myr) mimics those used by \cite{villaLarsen11}. The Padova isochrones used to determine the SFHs are not computed below an age of 4~Myr and are therefore not included in the SFHs of \cite{weisz11b}. However, the isochrones used to determine the cluster parameters do extend below 4~Myr and our formal cluster age dating extends to clusters of 1~Myr. Despite the differences in formal age limits, we expect the SFHs and the cluster ages to probe the same time bins.

We determine the approximate completeness limit of our sample by examining the $V$ magnitude histogram of all clusters. The peak in the number of clusters at the magnitude bin of 21.5 is used as the completeness limit. Our completeness limiting magnitude corresponds to $\sim$1500 and $\sim$3000 M$_{\odot}$ at the median distance of our sample for the two age ranges we study in this analysis of $<$10 and $<$100~Myr age ranges, respectively. Although the peak of clusters in the $V$ magnitude histogram is a rough indication of our completeness limit, we will show in Section 5.5 that the limit does not significantly change the overall results of our study. 

Figure 1 shows the colors of the clusters and the stochastic models over-plotted as a shaded region, where the ages of clusters and models are less then 100~Myr. Figure 2 shows the inferred age-mass diagram for all the clusters with an age less than 1 Gyr. The clusters with ages to the left of the vertical lines at log (age) of 7 and 8 constitute the cluster sample of the two age ranges ($<$10~Myr and $<$100~Myr) used in this study. It is interesting to note the lack of massive ($M \gtrsim 4.5 \times 10^4 M_{\odot}$) clusters in our selected age ranges compared to older ages.

\begin{figure}[h!]
  \begin{center}
  \includegraphics[scale=0.5]{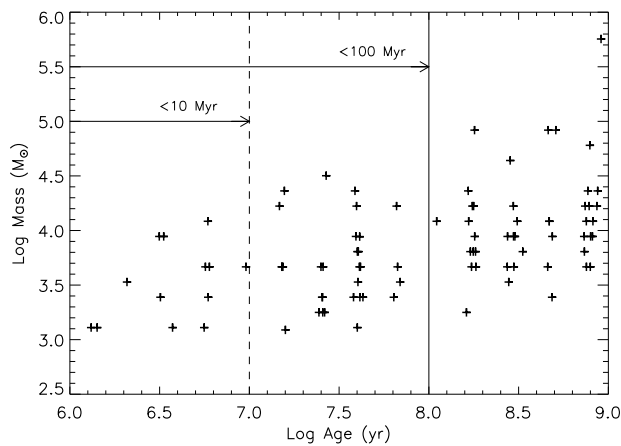}
  \caption{The age-mass diagram for all the clusters with an age less than 1 Gyr. The vertical lines at log (age) of 7 and 8 represent the upper end of the two age ranges used in this study. The horizontal arrows are visual representations of the two age ranges, $<$10 and $<$100~Myr.}
  \label{fig:figure2}
   \end{center}
\end{figure}  

Figure 3 shows the Cluster Mass Function (CMF) for the clusters with an age less then 100~Myr. We populate mass bins with an equal number of clusters (nine per bin) to avoid small number biases \citep{maiz05}. We also populate one bin with all of the clusters below the completeness limit; this bin is plotted as an open square and shows a dearth of clusters. The vertical error bars represent the Poisson statistics of the bin and the horizontal error bars show the mass range for each bin. The vertical dotted line corresponds to our completeness limit at the median distance of the galaxy sample. 

We fit the data above the completeness limit and find good agreement with multiple studies that show a nearly uniform power-law slope of $-$2 $\pm$ 0.2 \citep{zhang99,hunter03,bik03,degrijsb,mccrady07,fouesneau12}. The $\chi^2$ fit yields a slope of 1.94$\pm$0.26. We also fit a power-law to the data with a fixed slope of -2 to illustrate the difference between the fitted slope and the canonical slope. These slopes are graphically displayed in Figure 3 as solid and dashed lines for the fitted and fixed slopes, respectively. More recent studies show evidence that the CMF is more accurately represented as a Schechter function \citep{schechter76}, where the low cluster mass regime is characterized as a power-law and the high mass regime exponentially decays \citep[$M_{\star} > 10^5 M_{\odot}$;][]{gieles06b,Bastian08,gieles09b,larsen09,bastian12}. However, our single component power-law fit is justified since none of our clusters are near this high mass elbow in the Schechter function. 

\begin{figure}[h!]
  \begin{center}
  \includegraphics[scale=0.5]{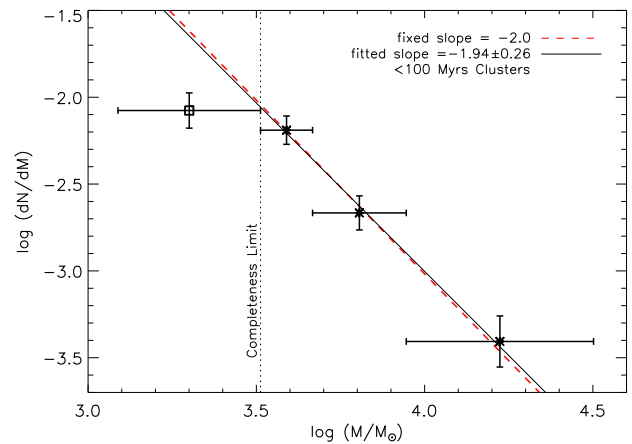}
  \caption{The Cluster Mass Function (CMF) of the cluster sample in the $<$100~Myr age range. The vertical error bars represent the Poisson error and the horizontal error bars show the mass range covered by each mass bin. The vertical dotted line represents the completeness mass at the median distance of the sample computed from the discrete models. Starting at the completeness limit, we populate equal number of cluster bins (9 per bin) to avoid small number biases. We also populate one bin below the completeness limit with all clusters below the completeness limit; this bin is plotted as an open square. The solid line is the fit to the data above the completeness limit. The dashed line is the fit with a fixed slope of -2 to facilitate a visual comparison between the canonical CMF slope and the unconstrained fit. The fit to the data shows agreement with a canonical CMF slope of -2.}
  \label{fig:figure4}
   \end{center}
\end{figure}  

\subsection{Star Formation Histories}
The Star Formation History (SFH) of each galaxy came from \cite{weisz11b} and was generated using MATCH \citep{dolphin02}. \cite{weisz11b} used a Salpeter IMF \citep{salpeter55} with the Padova stellar evolutionary models of \cite{marigo08}. These models are similar to those used for cluster fitting, but combines updated AGB tracks with the models of \cite{bertelli94} and \cite{girardi02}.

A full description of the method used to generate SFHs can be found in \cite{dolphin02}, however, we summarize the key points here. The SFH code allows the user to input an IMF, binary fraction, and allowable ranges in age, metallicity, distance, and extinction. Synthetic Color Magnitude Diagrams (CMDs) are created for every combination of the input parameters and compared to the observed CMD using a parametrized $\chi^2$. Minimizing the $\chi^2$ yields the most likely SFH of the galaxy.  Due to the presence of luminous main sequence and supergiant stars, the SFHs over the last 100~Myr considered can be derived very accurately \citep{dohm98,weisz08}.

\begin{center}
\begin{deluxetable*}{lcccccccc}[b]
\tabletypesize{\footnotesize}
\tablecolumns{9} 
\tablewidth{0pt} 
\tablecaption{Dwarf Galaxy Clustered Star Formation}
\tablehead{\colhead{Galaxy}         		&
	   					&
           \colhead{(4-10~Myr)}       		& 
	   					&
	   					&
           					&      		
           \colhead{(4-100~Myr)}       		&
	   					&
	   					\\ \hline
	   	 				&
	   \colhead{$\langle CFR \rangle_{10}$}       	&
	   \colhead{CMF Corr}       		&
	   \colhead{$\langle SFR \rangle_{10}$}    		&
	   \colhead{$M_V^{\rm brightest}$}\vline&
	   \colhead{$\langle CFR \rangle_{100}$}       	&
	   \colhead{CMF Corr}       		&
	   \colhead{$\langle SFR \rangle_{100}$}      	&
	   \colhead{$M_V^{\rm brightest}$}	\\
           	 				&
           \colhead{(M$_\odot$/yr)}      	&
	   \colhead{(factor)}			&
           \colhead{(M$_\odot$/yr)}      	&
           \colhead{(mag)}		\vline	&      		
           \colhead{(M$_\odot$/yr)}       	&
	   \colhead{(factor)}			&
           \colhead{(M$_\odot$/yr)}      	&
           \colhead{(mag)}			}
\startdata
ESO540-G030     &  \nodata   & \nodata & 0.0      &  \nodata           &  \nodata   & \nodata   & 2.28e-05 &  \nodata           \\ 
NGC404          &  \nodata   & \nodata & \nodata  &  \nodata           &  \nodata   & \nodata   & \nodata  &  \nodata           \\
KKH37           &  \nodata   & \nodata & 0.0      &  \nodata           &  \nodata   & \nodata   & 4.71e-04 &  \nodata           \\
NGC2366         & 4.07e-04   & 3.29    & 9.35e-02 & -8.52$\pm$0.02     & 1.73e-04   & 1.86      & 6.26e-02 & -8.52$\pm$0.02 \\
UGCA133         &  \nodata   & \nodata & 0.0      &  \nodata           &  \nodata   & \nodata   & 1.87e-05 &  \nodata   \\
UGC4305         & 3.01e-03   & 5.04    & 1.14e-01 & -8.88$\pm$0.01     & 7.51e-04   & 2.26      & 6.37e-02 & -8.88$\pm$0.01 \\
M81da           &  \nodata   & \nodata & 2.07e-03 &  \nodata           &  \nodata   & \nodata   & 1.66e-03 &  \nodata   \\
UGC4459         & 5.60e-04   & 7.94    & 3.25e-03 & -7.90$\pm$0.03     & 4.84e-05   & 5.75      & 6.53e-03 & -7.90$\pm$0.03 \\
UGC5139         &  \nodata   & \nodata & 1.75e-02 &  \nodata           &  \nodata   & \nodata   & 1.43e-02 &  \nodata           \\
FM2000\_1       &  \nodata   & \nodata & 3.59e-06 &  \nodata           &  \nodata   & \nodata   & 6.00e-06 &  \nodata           \\
BK3N            &  \nodata   & \nodata & 1.36e-05 &  \nodata           &  \nodata   & \nodata   & 1.57e-03 &  \nodata           \\
KDG61           &  \nodata   & \nodata & 4.71e-04 &  \nodata           &  \nodata   & \nodata   & 1.02e-04 &  \nodata           \\
arpsloop        &  \nodata   & \nodata & 0.0      &  \nodata           &  \nodata   & \nodata   & 2.54e-03 &  \nodata           \\
UGC5336         & 9.86e-04   & 2.76    & 3.71e-03 & -8.59$\pm$0.01     & 2.54e-04   & 2.01      & 2.69e-02 & -8.59$\pm$0.01 \\
LEDA166101      &  \nodata   & \nodata & 0.0      &  \nodata           &  \nodata   & \nodata   & 1.34e-05 &  \nodata           \\
UGC5428         &  \nodata   & \nodata & 0.0      &  \nodata           &  \nodata   & \nodata   & 0.0      &  \nodata           \\
UGC5442         &  \nodata   & \nodata & \nodata  &  \nodata           &  \nodata   & \nodata   & \nodata  &  \nodata           \\
IKN             &  \nodata   & \nodata & 0.0      &  \nodata           &  \nodata   & \nodata   & 1.64e-04 &  \nodata           \\
HS98\_117       &  \nodata   & \nodata & 1.42e-04 &  \nodata           &  \nodata   & \nodata   & 8.91e-06 &  \nodata           \\
DDO78           &  \nodata   & \nodata & 5.20e-06 &  \nodata           &  \nodata   & \nodata   & 2.18e-05 &  \nodata           \\
IC2574          & 4.10e-03   & 2.10    & 1.06e-01 & -9.12$\pm$0.03     & 1.29e-03   & 1.55      & 8.18e-02 & -9.12$\pm$0.03 \\
UGC5692         & 2.14e-04   & 5.00    & 7.34e-03 & -9.44$\pm$0.03     & 1.85e-05   & 3.62      & 1.28e-03 & -9.44$\pm$0.03 \\
KDG73           &  \nodata   & \nodata & 0.0      &  \nodata           &  \nodata   & \nodata   & 8.14e-04 &  \nodata           \\
NGC3741         &  \nodata   & \nodata & 1.44e-02 &  \nodata           &  \nodata   & \nodata   & 4.22e-03 &  \nodata           \\
NGC4163         &  \nodata   & \nodata & 9.93e-03 &  \nodata           &  \nodata   & \nodata   & 2.58e-03 &  \nodata           \\
UGCA276         &  \nodata   & \nodata & 0.0      &  \nodata           &  \nodata   & \nodata   & 1.98e-05 &  \nodata           \\
UGCA292         &  \nodata   & \nodata & 5.15e-03 &  \nodata           & 1.26e-04   & 2.82      & 3.35e-03 & -7.46$\pm$0.04 \\
UGC8091         &  \nodata   & \nodata & 1.65e-02 &  \nodata           &  \nodata   & \nodata   & 1.77e-03 &  \nodata           \\
UGC8201         &  \nodata   & \nodata & 3.16e-02 &  \nodata           & 1.29e-03   & 1.32      & 5.79e-02 & -9.16$\pm$0.02 \\
UGC8508         &  \nodata   & \nodata & 6.54e-03 &  \nodata           &  \nodata   & \nodata   & 2.77e-03 &  \nodata           \\
UGC8651         &  \nodata   & \nodata & 8.34e-04 &  \nodata           &  \nodata   & \nodata   & 3.41e-03 &  \nodata           \\
UGC8760         & 2.14e-04   & 5.00    & 3.03e-03 & -7.37$\pm$0.08     & 1.85e-05   & 3.62      & 4.29e-03 & -7.37$\pm$0.08 \\
UGC8833         &  \nodata   & \nodata & 2.52e-03 &  \nodata           &  \nodata   & \nodata   & 1.56e-03 &  \nodata           \\
KKR03           &  \nodata   & \nodata & 0.0      &  \nodata           &  \nodata   & \nodata   & 2.42e-04 &  \nodata           \\
UGC9128         &  \nodata   & \nodata & 2.79e-04 &  \nodata           & 1.77e-05   & 3.74      & 1.29e-03 & -5.45$\pm$0.12 \\
UGC9240         &  \nodata   & \nodata & 5.24e-03 &  \nodata           & 6.68e-05   & 4.44      & 8.86e-03 & -5.90$\pm$0.19 \\
KKH98           &  \nodata   & \nodata & 0.0      &  \nodata           &  \nodata   & \nodata   & 4.97e-04 &  \nodata           \\
\enddata
\tablecomments{The cluster formation and star formation results organized by two age ranges ($<$10~Myr and $<$100~Myr), as indicated by the top label. For each age range this table presents the Cluster Formation Rate (CFR), the Cluster Mass Function (CMF) correction used in Section 4, the CMD-based Star Formation Rate (SFR), and the $V$ absolute magnitude of the brightest cluster ($M_V^{\rm brightest}$). The CFRs include the IMF correction factor of 1.38 presented in Section 3.1. An ellipsis represents no data or no detectable clusters.}
\end{deluxetable*}
\end{center}
\vspace{-1cm}

\section{Cluster and Star Formation Properties}
In this section we describe the calculation the cluster formation efficiency ($\Gamma=CFR/SFR$) and star formation rate density ($\Sigma_{SFR}$). We then present the raw data and examine the $\Gamma$-$\Sigma_{SFR}$ and $M_V^{\rm brightest}$-SFR relations in our dataset. As we will show, no clear trend is seen in the $\Gamma$-$\Sigma_{SFR}$ relation, while the $M_V^{\rm brightest}$-SFR shows the expected correlation.  However, both plots show significant scatter whose source will be discussed in depth in Section 5.

\subsection{Cluster Formation Rates}
The cluster formation rate of each galaxy is computed by adding up the mass of clusters in each age range then dividing by the duration; these values are listed in Table 2. We have made a power-law CMF correction for each galaxy to account for missing faint clusters. The slope of the power-law is set to $-$2 and we have made the assumption that the least massive bound cluster would have $M \geq 100 $M$_\odot$. The assumption of a pure power-law CMF is valid due to the lack of any massive clusters and the agreement of the sample's CMF with the canonical $-$2 power-law (see Figure 3). The missing mass for galaxies with cluster detections are computed by integrating under the CMF from 100M$_\odot$ to the lowest mass bin where clusters are detected. Since we verified that each individual galaxy's CMF showed no significant deviation from a canonical CMF, we did not integrate to the completeness limit of each galaxy.

There are some dwarf galaxies in our sample that do not have any detectable clusters in the age ranges studied. These galaxies will be represented as an upper limit symbol (downward arrow). In the $\Gamma$-$\Sigma_{SFR}$ figures the upper limits correspond to the maximum total cluster mass undetectable and are calculated by integrating under a CMF with a slope of $-2$ between $100~M_{\odot}$ and the limiting completeness mass for each galaxy. The completeness mass is determined from our completeness magnitude limit (m$_V =$21.5). At the distance of each galaxy the completeness limiting magnitude is translated into a mass given the Bayesian model median mass for each age range. Since there are no clusters detected just above the completeness limit in galaxies without detected clusters, we assume that at most one cluster will be present at the completeness limit. This assumption translates into the amplitude of the CMF for galaxies without detected clusters. Dividing the missing mass under the CMF by the duration of each age range yields an upper limit $CFR$. Dividing the limiting $CFR$ by the measured $SFR$ produces an upper limit cluster formation efficiency ($\Gamma$). In the $M_V^{\rm brightest}$-SFR figures the upper limits correspond to our completeness $V$ magnitude of 21.5 at the distance of each galaxy.

We determine the cluster formation rates in two different age ranges, $<$10~Myr and $<$100~Myr.  The cluster formation rates in the $<$10~Myr age range are based on small numbers of clusters and also suffers from uncertainties in the age. However, we have verified that each of these clusters show moderate to strong H$\alpha$ emission which traces star formation of the appropriate age range (see Section 3.3). The $<$100~Myr age range clusters will have larger number statistics due to the longer duration, but here the effects of cluster destruction may be important.

An IMF correction needs to be applied to our cluster sample due to the different adopted IMFs assumed for cluster parameter (Kroupa IMF) and the SFR (Salpeter IMF) determinations. A Kroupa IMF deviates from a Salpeter IMF to lower numbers of stars at low masses and therefore underestimates the mass in clusters if the CFR and SFR are compared. Integrating under both IMFs normalized to produce the same number of stars yields a mass ratio of 1.38. We multiply our CFRs by this factor to account for the different IMF assumptions.

\subsection{Star Formation Rates}
We calculate the field SFR from the CMD-based SFHs (e.g., the SFR as a function of time and metallicity) of \cite{weisz11b}. The SFHs were generated with the method described in Section 2.5. The corresponding SFRs are calculated by taking the total stellar mass formed during the age range and divided by the duration. The resulting average SFRs are listed in Table 2. It should be noted that there may be periods of higher or lower SFRs during the adopted time interval.

\subsection{$\Sigma_{SFR}$}
The star formation rate surface density ($\Sigma_{SFR}$) is calculated by normalizing the SFR by the area of the galaxy being studied: typically the whole galaxy in our sample. Most of our galaxies have a small enough angular size to fit entirely within HST ACS camera field of view. The size of these galaxies were determined using the area defined by the 3.6 $\mu$m ellipses of \cite{dale09}, listed in Table 1. Since FUV and H$\alpha$ flux are SFR indicators of 100~Myr and 10~Myr star formation, respectively, we have visually verified that these ellipses contain both the FUV and H$\alpha$ emission in each galaxy. The FUV and H$\alpha$ images were taken as part of the 11~Mpc H$\alpha$ UV Galaxy Survey \citep[11HUGS;][]{kennicutt08,lee09a} In five galaxies, the 3.6 $\mu$m ellipses were larger than the ACS FOV. The normalizing area for these galaxies were taken as the number of pointings times the ACS FOV ($202^{\prime\prime} \times 202^{\prime\prime}$).

\subsection{$\Gamma$-$\Sigma_{SFR}$}
With both the cluster formation rates and the star formation rates calculated, the cluster formation efficiency can be quantified. By definition $\Gamma$ is the ratio of the CFR divided by the SFR. 

Figure 4 shows the $\Gamma$-$\Sigma_{SFR}$ plot. The $<$10~Myr data show a large scatter to higher $\Gamma$s, whereas the $<$100~Myr data show lower $\Gamma$ values. The downward arrows are galaxies where no clusters were detected and represent upper limits on the cluster formation efficiency for these galaxies. Low mass clusters below our completeness limit may be present, but the total cluster mass will fall below the upper limit arrows. 

Due to the small numbers of clusters in each of our dwarf galaxies, we expect the scatter seen in our data to be dominated by stochastic effects. The propagation of formal errors in cluster age and mass does not include this scatter. In Section 5 we model the stochastic effects and provide resulting confidence intervals for each galaxy.

\begin{figure}[h!]
  \begin{center}
  \includegraphics[scale=0.5]{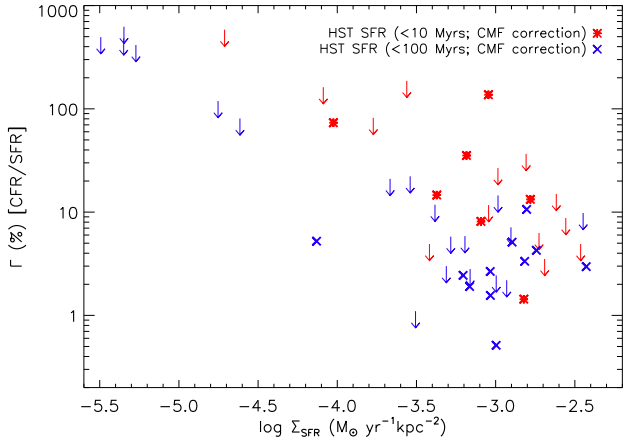}
  \caption{The cluster formation efficiency ($\Gamma$=CFR/SFR) versus the star formation rate density ($\Sigma_{SFR}$=SFR/kpc$^2$) for two age ranges. The data have been corrected for incompleteness via a CMF with a slope of -2. Data from the $<$10~Myr interval are presented as red asterisks and the $<$100 interval as blue X's. The downward arrows are galaxies where no clusters were detected and represent upper limits on the total mass in clusters; the upper limits have the same color coding as the non-upper limit data points. The $<$100~Myr age data show up to a factor of 100 variation in the cluster formation efficiency at fixed star formation rate density. A large amount of this scatter is toward low cluster formation rates.}
  \label{fig:figure3}
   \end{center}
\end{figure}  

\subsection{$M_V^{\rm brightest}$-SFR}
The scatter in $\Gamma$ is motivation for analyzing another parameter space that compares cluster formation and star formation, namely the $V-band$ luminosity of the brightest cluster ($M_V^{\rm brightest}$) versus the star formation rate. The brightest cluster in a galaxy has previously been found to correlate with the galaxy's SFR, as a larger SFR results in a larger cluster population sampling higher mass clusters \citep{larsen02}. In Figure 5 we plot the brightest cluster in the appropriate age range ($<$10~Myr or $<$100~Myr) versus the galaxy's SFR. Both age ranges show a positive trend with a moderate amount of scatter. The upper limit data points are galaxies with no detectable clusters and represent the absolute $V$ magnitude of our completeness limit at the distance of each galaxy. At the higher SFR end of our sample, these upper limits are often 3--4 magnitudes fainter than galaxies with detected brightest clusters and similar SFRs. Stochasticity in the number of rare, massive clusters may contribute significantly to the scatter; we model the stochastic effects on this relationship in Section 5.

\begin{figure}[h!]
  \begin{center}
  \includegraphics[scale=0.5]{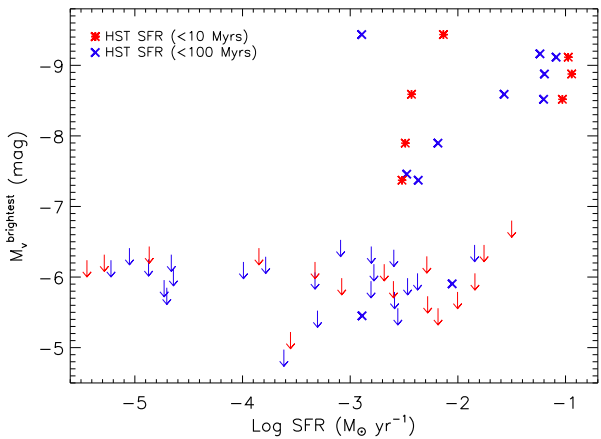}
  \caption{The $M_V^{\rm brightest}$-SFR for the two age ranges. Data from the $<$10~Myr interval are presented as red asterisks and the $<$100 interval as the blue X's. The downward arrows are galaxies where no clusters were detected and represent upper limits on the total mass in clusters; the upper limits have the same color coding as the non-upper limit data points. Both age range's brightest clusters show a positive trend with SFR.}
  \label{fig:figure5}
   \end{center}
\end{figure}  


\section{Results}
In this section we discuss the results of previous studies on the relation between star formation and cluster formation in higher mass galaxies. We take these studies' results as presented in the original work, highlighting sources of uncertainty and differences from out analysis. We then present a comparison of our $\Gamma$-$\Sigma_{SFR}$ and M$_V^{\rm brightest}$-SFR relations to previous work done in higher mass galaxies. In Section 5 we present a detailed analysis of the scatter in these relationships using simulations of stochastic cluster formation.

\subsection{Previous Studies}
All of the previous studies of cluster formation efficiency share several common elements.  In each, cluster masses were determined using continuous models, as opposed to the discrete models used in our analysis.  Also, each uses a cluster mass function correction assuming a $-$2 slope power-law at the low mass end. Both \cite{goddard10} and \cite{adamo11} used SFRs derived from integrated fluxes, whereas \cite{villaLarsen11} used CMD-based SFRs derived from a similar method as the one presented in this study. The integrated flux SFRs were based on the Kennicutt-Schmidt law which assumes a Salpeter IMF \citep{kennicutt98}. To achieve consistency between the IMFs used in estimating the cluster formation rate, the cluster masses from this study, \cite{goddard10}, and \cite{adamo11} needed to be multiplied by a factor of 1.38.  \cite{villaLarsen11} used the same IMF in determining both quantities. The characteristic timescales for these studies are $<$10~Myr for \cite{goddard10} and \cite{adamo11}, and $<$100~Myr for \cite{villaLarsen11}.

Additional efforts were made by \cite{villaLarsen11} to investigate the effects of cluster destruction on $\Gamma$ by modeling the destruction-dependent luminosity function. However, we treat the $\Gamma$ versus $\Sigma_{SFR}$ relationship as purely observational and therefore only use their values labeled P1 \citep[Paper I analysis method;][]{villaLarsen10}. The P1 values are those obtained without cluster destruction taken into account. The results of all three studies are listed in Table 3.

\begin{center}
\begin{deluxetable}{lccccc}[h!]
\tabletypesize{\footnotesize}
\tablecolumns{6} 
\tablewidth{0pt} 
\tablecaption{Other Studies Results}
\tablehead{\colhead{Galaxy}        		&
	   \colhead{CFR}       			&
	   \colhead{SFR}       			&
	   \colhead{Area} 			&
	   \colhead{$\Gamma$}  			&
	   \colhead{Ref.}       		\\
           					&
           \colhead{(M$_\odot$/yr)}       	&
           \colhead{(M$_\odot$/yr)}       	&
           \colhead{(kpc$^2$)}     	  	&
	   \colhead{($\%$)}			&
       						}
\startdata
NGC1569        & 0.05  &0.3626 & 13.0  & 13.9$\pm$0.8          & \tablenotemark{a} \\
NGC3256        & 10.57 &46.17  & 74.9  & 22.9$^{+7.3}_{-9.8}$  & \tablenotemark{a} \\
NGC5236        & 0.103 &0.3867 &0.708  & 26.7$^{+5.3}_{-4.0}$  & \tablenotemark{a} \\
NGC6946        & 0.022 &0.1725 & 37.5  & 12.5$^{+1.8}_{-2.5}$  & \tablenotemark{a} \\
LMC		& 0.007	&0.1201 & 79.0 	& 5.8$\pm$0.5 		& \tablenotemark{a} \\
SMC		& 0.002	&0.0426 & 58.9  & 4.2$^{+0.2}_{-0.3}$ 	& \tablenotemark{a,d} \\
Milky Way	& 0.011	&0.1508 & 12.6  & 7.0$^{+7.0}_{-3.0}$ 	& \tablenotemark{a} \\
ESO338-IG04	& 1.6	& 3.2	& 2.07	& 50$\pm$10		& \tablenotemark{b} \\
Haro 11		& 11.2	& 22.0	& 10.2	& 50$^{+13}_{-15}$	& \tablenotemark{b} \\
ESO185-IG13	& 1.7	& 6.4	& 12.3	& 26$\pm$5		& \tablenotemark{b} \\
Mrk 930		& 1.33	& 5.34	& 9.05	& 25$\pm$10		& \tablenotemark{b} \\
SBS0335-052E	& 0.64	& 1.3	& 1.37	& 49$\pm$15		& \tablenotemark{b} \\
NGC5236        & 0.038 & 0.39  & 28.7  & 9.8                   & \tablenotemark{c} \\
NGC7793        & 0.015 & 0.15  & 23.1  & 9.8                   & \tablenotemark{c} \\
NGC1313        & 0.061 & 0.68  & 60.0  & 9.0                   & \tablenotemark{c} \\
NGC0045        & 0.009 & 0.05  & 49.0  & 17.3                  & \tablenotemark{c} \\
NGC4395        & 0.005 & 0.17  & 36.5  & 2.6                   & \tablenotemark{c} \\
\enddata
\tablecomments{The results from previous studies presented here are from: \tablenotemark{a}{\cite{goddard10}}, \tablenotemark{b}{\cite{adamo11}}, and \tablenotemark{c}{\cite{villaLarsen11}}, and \tablenotemark{d}{\cite{gielesBastian08}}. The quantities listed are the galaxy name, Cluster Formation Rate (CFR), the Star Formation Rate (SFR), the normalizing area, and $\Gamma$ (CFR/SFR). The \cite{goddard10} and \cite{adamo11} values have the IMF correction factor of 1.38 incorporated into them.}
\end{deluxetable}
\end{center}

\subsection{$\Gamma$-$\Sigma_{SFR}$}
We combine our $\Gamma$-$\Sigma_{SFR}$ measurements with data from the studies of \cite{goddard10}, \cite{villaLarsen11}, and \cite{adamo11} in Figure 6. The solid line is the linear regression fit to the data of \cite{goddard10}. The dotted lines above and below the trend line are three times the standard deviation of all the previous study's data around the trend line. Although, the relationship of \cite{goddard10} is supported by many sources, none of them are probing the star formation rate regime of our data. 

\begin{figure}[h!]
  \begin{center}
  \includegraphics[scale=0.5]{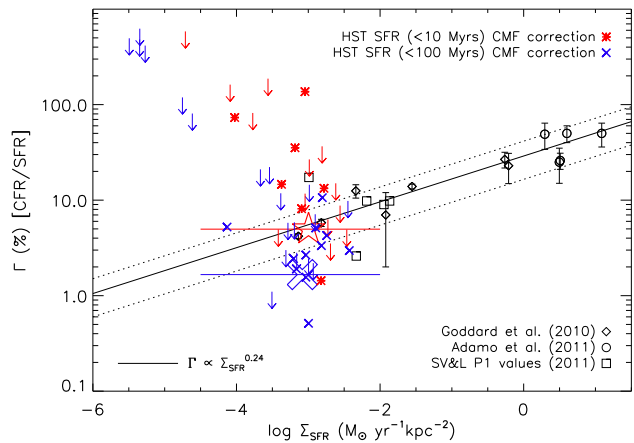}
  \caption{The $\Gamma$-$\Sigma_{SFR}$ plot, similar to Figure 4, but with the previous study's data over-plotted as open symbols. The solid line is the trend from \cite{goddard10} and the dotted lines are the 3 times the standard deviations of all the previous data. Data from both age bins show poor agreement with the trend of previous studies. The majority of the galaxies in the $<$100~Myr age range fall significantly below the trend indicating a dearth of total cluster mass. To study the effects of low number statistics we have binned our data for the two age ranges. The $<$10~Myr and $<$100~Myr binned data points are the large, red, open star and large, blue, open X, respectively; the x-axis error bars represent the width of the bin. The binned $<$10~Myr data show good agreement with the expected trend, but the $<$100~Myr data show a dearth of clusters despite the better number statistics of the longer age range.}
  \label{fig:figure6}
   \end{center}
\end{figure}  

We find poor agreement for both age ranges with the extrapolation of the \cite{goddard10} relationship. The $<$10~Myr age range data are discrepant to both high and low $\Gamma$ values whereas the majority of the $<$100~Myr age range galaxies show a dearth of cluster formation. UGC5692 deviates above the trend line in both age ranges and NGC2366 deviates below the line in both age ranges. This scatter suggests there may be a true variation in the cluster formation properties of these galaxies. We discuss the outliers in more detail in Section 5.

To examine the effects of low number statistics we have binned all of the clusters into one cluster population. We performed an analysis of the combined cluster population by treating these clusters as if they occupy a single galaxy.  The procedure is the same as described in Section 3.1. We calculate the total cluster mass in the $<$10 Myr and $<$100 Myr age ranges, including all galaxies with log($\Sigma_{SFR}$) values $> -4.5$.  The binned $\Gamma$ values are calculated by dividing a CMF corrected cluster formation rate by the sum of the star formation rates; including the upper limit star formation rates. The $\Sigma_{SFR}$ values are determined by summing the SFRs and dividing by the sum of the areas.  The binned $<$10 and $<$100 Myr data are shown on Figure 6 as a large, red, open star and a large, blue, open X, respectively.

The $<$10~Myr data show good agreement with the expected trend of \cite{goddard10}. However, the binned $<$100~Myr data shows a significant deviation to lower $\Gamma$ values (lower total cluster mass). Due to its longer age-range, the $<$100~Myr age range has more clusters and better number statistics than the $<$10~Myr data. However, it contains fewer clusters than expected from the $\Gamma$-$\Sigma_{SFR}$ relationship. We will discuss possible sources of the dearth of clusters in the $<$100 Myr age range in Section 5.

\subsection{$M_V^{\rm brightest}$-SFR}
Plotting the brightest cluster versus the galaxy-wide SFR facilitates the analysis of clustered star formation by directly testing stochastic effects. \cite{larsen02} originally proposed the statistical ``size-of-sample" effect explanation for this correlation, such that higher SFR events can sample higher into the CMF to create higher mass (assumed brighter) clusters. In other words, this relationship holds because of stochastic effects. Figure 7 shows the data from this study over-plotted onto those of \cite{larsen02}, \cite{Bastian08}, and \cite{adamo11}. The trend line is a linear regression fit of the data from \cite{larsen02} performed by \cite{weidner04}. The dotted lines above and below the trend represent three times the standard deviation of all previous data points. 

\begin{figure}[h!]
  \begin{center}
  \includegraphics[scale=0.5]{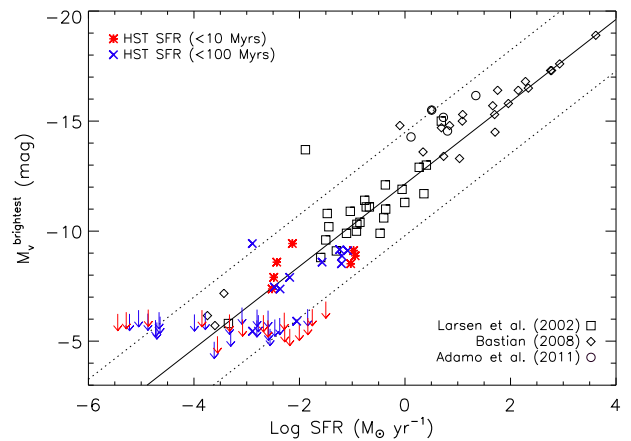}
  \caption{The $M_V^{\rm brightest}$-SFR for the two age ranges. This is similar to Figure 5, but with the previous data over-plotted as open symbols. The solid line is the linear regression fit to the data of \cite{larsen02} performed by \cite{weidner04}. The dotted lines are the 3 times the standard deviations of all the previous data. Both age ranges show good agreement with the established trend, but there are a few outliers. Upper limits (no detectable clusters) below the trend line are significant since these galaxies should have formed a detectable bright cluster, given the star formation rate.}
  \label{fig:figure7}
   \end{center}
\end{figure}  

There are outliers falling both above and below the expected $M_V^{\rm brightest}$-SFR relation. The typical $V$ magnitude error for data presented in Figure 7 is approximately 0.03 magnitudes. The five upper limit low outliers near a log {\it SFR}(M$_\odot~{\rm yr}^{-1}$) of $-2$ belong to the $<$10~Myr age range. These upper limits are significant because the brightest clusters predicted by the trend are brighter than the completeness limits of these galaxies, but no clusters of these magnitudes are detected. Conversely, the $<$100~Myr data show no significant outliers below the trend. It is interesting to note that all of the $\Gamma$-$\Sigma_{SFR}$ low outliers show good to moderate agreement with the $M_V^{\rm brightest}$-SFR trend.


There is also one high outlier galaxy (UGC5692) in the $<$100~Myr age range that lies significantly above the trend line. This galaxy is also a high outlier in the $\Gamma$-$\Sigma_{SFR}$ relationship. However, \cite{Bastian08} showed that outliers above the trend were in fact not outliers after using the youngest ($t_{\rm{age}} \lesssim 10~$Myr) clusters to show that the brightest cluster is more likely to be young and therefore more accurately reflects the recent SFR. This is true for UGC5692, whose brightest cluster has an age less than 10~Myr. UGC5692 has undergone a recent increase in SFR; using the $<$10~Myr data point (log {\it SFR}(M$_\odot~{\rm yr}^{-1}$)$\sim-3$ and the same magnitude) brings this galaxy within the scatter of the relationship. However, the lower limits at high SFRs in the $<$10~Myr sample do seem to violate the $M_V^{\rm brightest}$-SFR correlation proposed by \cite{Bastian08}. A full discussion of these outliers be presented below.

\section{Discussion: Is the Scatter Due to Stochasticity}
The previous section showed that clustered star formation in dwarf galaxies does not necessarily agree with established cluster-host relationships. The sample shows a significant amount of scatter around these relationships with a large number of significant outliers, particularly in the $\Gamma$-$\Sigma_{SFR}$ relationship. However, because the overall SFRs of our dwarf galaxy sample are lower (due to both smaller galaxy mass and overall star formation efficiency), the cluster population is much more likely to be affected by stochastic sampling of the cluster mass function. Therefore we examine the effect of stochastic sampling of the CMF, and whether this sampling can account for the observed scatter in both the $\Gamma$-$\Sigma_{SFR}$ and M$_V^{\rm brightest}$-SFR relations. 

\subsection{SLUG}
We utilize the publicly available code ``SLUG" \citep[Statistically Lighting Up Galaxies;][]{slug1,slug2} to model these effects. SLUG creates a simulated galaxy by generating clusters and stars via stochastic sampling of both the CMF and the stellar IMF, respectively. The code allows the user to turn on and off cluster destruction and to set the fraction of stars that form in clusters. 

In addition to providing a tool to incorporate the variability in our data due to sampling of small numbers of clusters, the SLUG simulations also allow us to directly compare our simulations to data without making a CMF correction for clusters fainter than our completeness limit.  Half of the 17 galaxies with detected clusters have fewer than three clusters.  Consequently, the CMF correction for these galaxies is highly uncertain as a result of fitting the amplitude to one or two data points.  The direct comparison enabled by SLUG eliminates this correction, and incorporates the uncertainty in this mass correction into our stochasticity estimates.


\subsection{$\Gamma$-$\Sigma_{SFR}$}
To evaluate the scatter in $\Gamma$ we use SLUG to simulate only the cluster population of each galaxy.  The low values of $\Gamma$ of our galaxies mean that the amount of mass in clusters is a small fraction of the total star formation in the galaxies.  Therefore, we expect the stochasticity of the cluster formation process to dominate the scatter in $\Gamma$, and simulate only the effects of stochasticity in the cluster population. 

For our simulations, we assume that a universal cluster formation mechanism is accurately described by the \cite{goddard10} relationship. We input into SLUG the CFR expected if each galaxy fell on the $\Gamma$ versus $\Sigma_{SFR}$ relationship found by \cite{goddard10}, given the galaxy's measured $\Sigma_{SFR}$. SLUG allows the user to set the fraction of stars that form in clusters via a parameter with values that range from $0\leq f \leq 1$. Simulations can generate stars without clusters ($f$=0), clusters without stars ($f$=1), or some combination of both. Since we only want SLUG to simulate the cluster population of a galaxy, we input a CFR instead of the galaxy's SFR, we force all stars to form in clusters, and we disable cluster destruction. The combination of these input parameters effectively creates a cluster population based on the mass in clusters expected from the cluster formation rate observed in each galaxy. 

For each galaxy, we run 1000 simulations where each iteration randomly generates a new cluster population (ignoring field stars) from a CMF with a slope of $-$2 and a Salpeter IMF is assumed to populate each cluster. Our simulations also include the same stellar isochrones as those used for cluster fitting. The mass of clusters generated with a luminosity brighter than our completeness limit is added up and divided by the measured SFR to produce a simulated $\Gamma$. The median of the simulations is used as the prediction of the $\Gamma - \Sigma_{SFR}$ simulations, while the middle 68.3\% of the simulations are taken as the 1$\sigma$ confidence interval. To make a direct comparison between the SLUG simulations and our observed cluster sample, we have also cut any real clusters fainter than our completeness limit when comparing our data to the simulations.

While the absolute $\Gamma$ values of the simulations are based on the \cite{goddard10} relationship, the scatter seen in the data points should generically reflect how much of the scatter in our data results from cluster stochasticity regardless of the relationship between $\Gamma$ and $\Sigma_{SFR}$.

The results for the $<$10~Myr and $<$100~Myr age ranges are graphically shown in Figure 8, in the top and bottom panels, respectively. Each galaxy is evenly spaced on the $x$-axis, which has been sorted by $\Sigma_{SFR}$ with increasing values to the right. The open triangles are the SLUG simulated galaxies, and the red asterisks and blue x's are the $<$10~Myr and $<$100~Myr age range data points with detected clusters above our completeness limit, respectively. The simulated $\Gamma$ values are set to a value of 0 when the median of the simulations produced no clusters. Simulations for the majority of the lowest SFR galaxies (left of UGC5139 for the both age ranges) produced very few detectable clusters in agreement with the lack of clusters observed in those galaxies.

There are three major outliers (UGC5336, UGC8760, and UGC4459) to high $\Gamma$ values in the $<$10~Myr age range, and one (UGC5692) in the $<$100~Myr age range. These galaxies fall noticeably above the simulations. None of the simulations for UGC5336 nor UGC4459 and less than 1\% and 7\% of the simulations for UGC8760 and UGC5692, respectively, produced $\Gamma$ values as high as the observed data points. This makes UGC5692 a $\sim2\sigma$ outlier and UGC8760 a $\sim3\sigma$ outlier. Since 1000 simulations were run, the lack of overlap with UGC5336 and UGC4459 make these galaxies $>3\sigma$ outliers. These galaxies provide strong evidence that more clusters are formed than expected from $\Gamma$-$\Sigma_{SFR}$ relationship. 

\begin{figure}[h!]
  \begin{center}
  \includegraphics[scale=0.5]{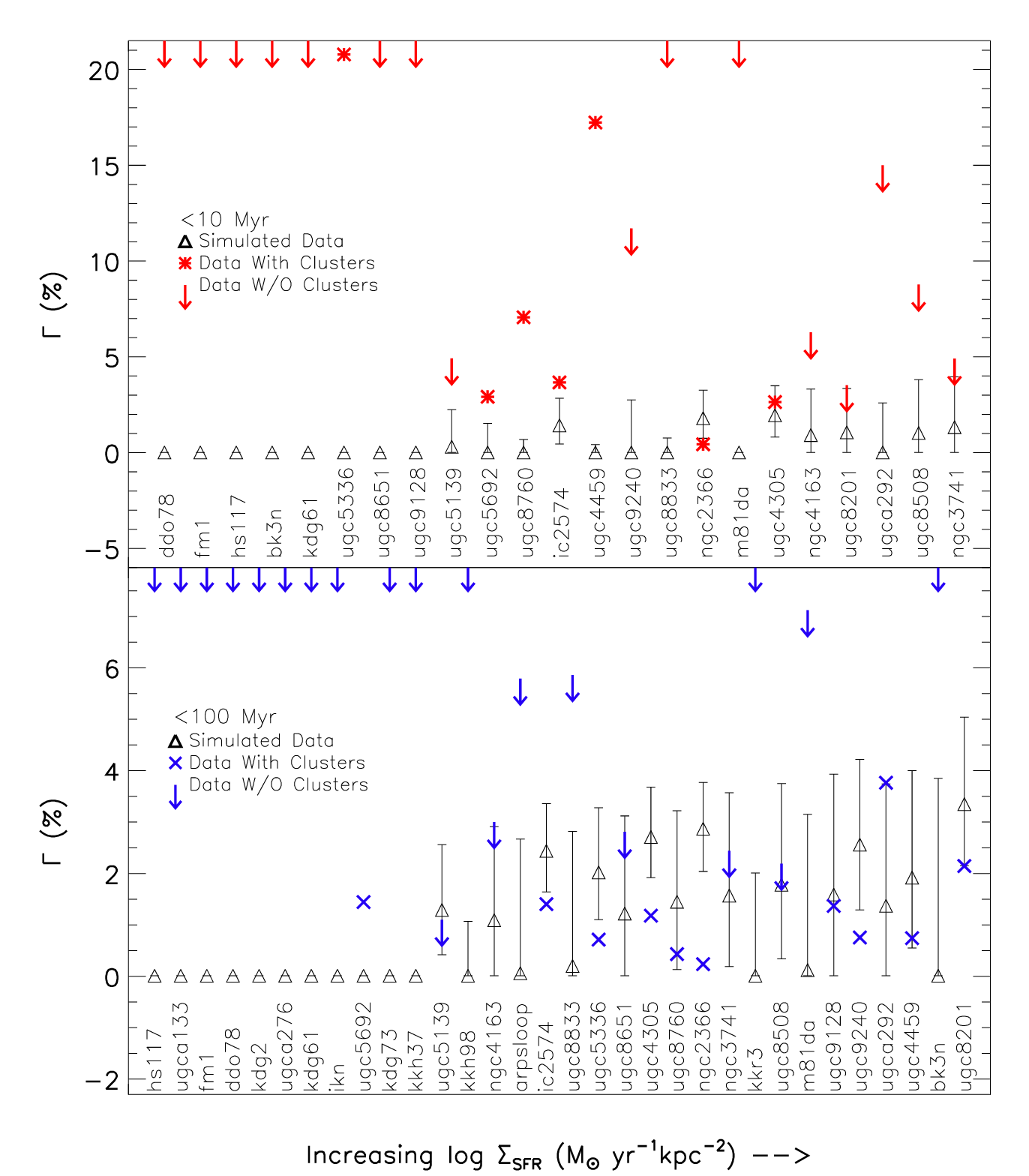}
  \caption{The simulated $\Gamma$ results from SLUG, where the top and bottom panels are the $<$10~Myr and $<$100~Myr age ranges, respectively. Each galaxy is presented on the $x$-axis evenly spaced and sorted by $\Sigma_{SFR}$, with increasing values to the right. The open triangles are the SLUG simulated galaxies and the error bars are the 68.3\% confidence inertvals of the 1000 realizations of the simulations. The blue X's are the observed data \emph{without} CMF corrections and the downward arrows are the maximum total cluster mass for galaxies with no detectable clusters. The majority of the SLUG simulations in both age ranges agree with the observed data since they are within at least 2 sigma. However, there are a few significant outliers above and below the simulations which might indicate a true variation in the cluster formation of these galalxies.  The $<$100~Myr age range simulations systematically fall above the observed data. This discrepency suggests that stochasticity cannot account for all of the scatter to low cluster formation seen in the $<$100 interval data.}
  \label{fig:figure8}
   \end{center}
\end{figure}  

There is one galaxy that falls significantly below the $\Gamma$-$\Sigma_{SFR}$ trend line in both age ranges: NGC2366. This galaxy also falls nearly $3\sigma$ below the SLUG simulations in the $<$100~Myr age range as well. \cite{billett02} also found this galaxy to have an unusual dearth of cluster formation compared to other galaxies with similar star formation properties. NGC2366 has spatially extended pockets of star formation where a lack of cluster formation could be due to a lower gas density. The gas density of dwarf irregular galaxies at intermediate and extended radii is typically well below the threshold of large-scale star formation \citep{hunter98,leroy08}. Also, there are large blue, nebulous active star forming regions in this galaxy that could hide young massive star clusters which may not yet be detectable. However, we consider hidden clusters an unlikely explanation since many other galaxies in our sample (e.g., IC2574, UGC4305, and UGC4459) also contain large nebulous regions and all either have good agreement with or lie above the $\Gamma$-$\Sigma_{SFR}$ trend in at least one age range. This implies that NGC2366 has truly formed fewer clusters than predicted by the $\Gamma$-$\Sigma_{SFR}$ relationship. 

We have examined the clusters in the outlier galaxies carefully to insure that their discrepant $\Gamma$ values are not due to erroneous age and mass determinations. Since our galaxies are close enough to partially resolve the stars in each cluster, we can broadly verify the age of the clusters in these galaxies. A young CMD will have a well defined main sequence and an old cluster will have a well defined red giant branch. All of the clusters in these outlier galaxies exhibit properties of a young stellar population of stars within the them. Furthermore, we have double checked the discrepant $<$10~Myr clusters' ages by confirming moderate to strong H$\alpha$ emission (SFR indicator of age less than 10~Myr) in each of these clusters and little to no H$\alpha$ emission in older clusters. The high and low outliers strongly suggest that the cluster formation efficiency shows significant variation in dwarf galaxies.

To test how well our data is described by the SLUG simulation model for the collection of galaxies as a whole, we calculate a reduced $\chi^2$ for each age range. We assign the observed data points as the expected value and treat the simulations as the ``measured" data points with errors. We then evaluate the goodness of the fit by calculating the reduced $\chi^2$ between the simulations and the observed data:

\begin{equation}
\chi_{\rm red}^2 = \frac{1}{N-p} \sum_{i=1}^N \frac{(\Gamma_{\rm sim} - \Gamma_{\rm obs})^2}{\sigma_{\rm sim}^2},
\end{equation}

\noindent where \emph{N} is the degrees of freedom (number of galaxies), \emph{p} is number of parameters being fit (one for the $\Gamma$ values), $\Gamma_{\rm sim}$ is the median simulated data point, $\Gamma_{\rm obs}$ is the observed data point, and $\sigma_{\rm sim}$ is the confidence interval of the simulations.  Upper limits are included in the estimate of $\chi^2_{\rm red}$ only when they fall below the median of the simulations. The $<$10~Myr and $<$100~Myr $\chi^2_{\rm red}$ values are 3.6 and 7.2, respectively.

Large $\chi^2_{\rm red}$ values could result from: (1) real deviation of the cluster formation efficiency from the $\Gamma-\Sigma_{SFR}$ relationship, (2) cluster destruction; (3) non-Gaussian stochastic errors as determined from SLUG; or (4) additional errors in our data or analysis. Number four includes errors in the mass and age determination, the cluster identification, and the completeness limit. 

Most of the simulations in the $<$100~Myr age range show moderate agreement on a galaxy-by-galaxy basis with the observed data since they are within the $2\sigma$ spread of the SLUG simulations. However, the dwarf galaxy sample as a whole shows a systematic offset to lower $\Gamma$ values compared to the stochastic simulations. The moderate agreement of most galaxies with the SLUG simulations in this age range imply that stochastic effects account for most of the scatter in $\Gamma$, but the systematic offset of the sample as a whole to lower $\Gamma$ values and the higher relative reduced $\chi^2$ value of the $<$100~Myr, compared to the $<$10~Myr age range, implies the presence of some additional source of scatter.

One explanation for the lack of cluster formation in the $<$100~Myr age range, compared to the $<$10~Myr age range, is cluster destruction. The primary mechanism responsible for cluster destruction is currently under debate (see Section 5.3), but its effect should be significant in a longer age range ($<$100~Myr) compared to a shorter age range ($<$10~Myr). This effect is indeed observed in our data. Seven galaxies with cluster detections and four without detections (upper limits) in the $<$100~Myr age range show significant deviations from the $\Gamma$-$\Sigma_{SFR}$ relationship to lower $\Gamma$ values (i.e. low total cluster mass), whereas only one galaxy with a cluster detection and one upper limit data point in the $<$10~Myr age range show a low $\Gamma$ deviation (see Figure 6). In other words, 11 data points in the $<$100~Myr age range are discrepant compared to only 2 in the $<$10~Myr age range. Furthermore, the improved number statistics via binning the entire cluster sample shows that the binned $<$100~Myr data are also discrepent to lower $\Gamma$ values compared to the binned $<$10~Myr data (see large, open symbols in Figure 6). It is likely that some form of cluster destruction is responsible for at least some of the additional scatter in our sample.

\subsection{Constraints on Cluster Destruction}
More scatter towards low cluster mass is seen in the $\Gamma-\Sigma_{SFR}$ relationship for the $<$100~Myr age range that cannot be fully accounted for by stochasticity. If cluster destruction is responsible for this deficit, then the $M_V^{\rm brightest}$-SFR relationship will provide evidence for whether or not the most massive clusters are surviving. There are two popular models of cluster destruction, (1) mass-independent destruction and (2) mass-dependent destruction. The mass-independent model is dominated by internal mechanisms, such as infant mortality, and does not preferentially destroy a cluster based on its' mass \citep{hills80,fall05,fall09,whitmore07,whitmore10,chandar06,chandar10a,chandar10b}. The mass-dependent model, on the other hand, is dominated by external tidal forces such as proximity to giant molecular clouds or interior sections of spiral galaxies, and thus preferentially destroys smaller mass clusters \citep{spitzer58,henon61,baumgardt03,lamers05,gieles07,bastian11}. Unfortunately, $\Gamma$ does not distinguish which clusters are missing from the total mass of clustered star formation and therefore cannot distinguish between these models. However, the $M_V^{\rm brightest}$-SFR relation correlates the brightest (and therefore most massive) cluster to the SFR of the galaxy, and analysis of both relationships may provide evidence of a dominant cluster destruction model. 

The $\Gamma$-$\Sigma_{SFR}$ relationship is sensitive to both cluster destruction models since both effect the overall cluster mass. However, the $M_V^{\rm brightest}$-SFR relationship should be unaffected by a mass-dependent destruction model and should show no major deviation from the trend since a mass-dependent cluster destruction model predicts that massive clusters will preferentially survive. If a mass-dependent model is dominant then the low $\Gamma$ galaxies should show agreement with the $M_V^{\rm brightest}$-SFR relationship. This cross-relationship prediction is seen in our data. Of the 11 galaxies with significantly low $\Gamma$ values, 4 out of 4 upper limits and 7 out of 7 galaxies with cluster detections show good to moderate agreement with $M_V^{\rm brightest}$-SFR relation. This suggests that the missing mass in the cluster formation seen in the $<$100~Myr age range is not due to random destruction, but rather mass-dependent cluster destruction where the high mass (brightest) clusters preferentially survive. This is evidence for mass-dependent cluster destruction in our dwarf galaxy sample. It is also interesting to note that the increasing upper envelope in the age-mass distribution of Figure 2 (e.g., older clusters tend to be more massive) is a ``size-of-sample" effect. This effect is evidence that significant mass-independent destruction is not occurring in our data \citep{gielesBastian08}.

\subsection{$M_V^{\rm brightest}-SFR$}
We again use SLUG to test the effects of stochasticity on the results of the $M_V^{\rm brightest}-SFR$ diagram (Figure 7). The same simulations as those in Section 5.1 are used, but this time we extract the brightest cluster formed in each iteration. The median brightest cluster is set as the simulation data point and the middle 68.3\% of the simulations is set as the 1$\sigma$ confidence interval.

Figure 9 shows the $<$10~Myr and $<$100~Myr age ranges in the upper and lower panels, respectively. The open triangles are the SLUG simulated galaxies, and the red asterisks and blue x's are the $<$10~Myr and $<$100~Myr age range data points, respectively, with detected clusters above our completeness limit. The simulated $M_V^{\rm brightest}$ values are set to a value of 0 when the median of the simulations produced no clusters. The $x$-axis has been sorted by SFR and is increasing to the right. Simulations for the majority of the lowest SFR galaxies (left of UGC8833 and KKR3 for the $<$10~Myr and $<$100~Myr age ranges, respectively) produced very few detectable clusters in agreement with the lack of bright clusters observed in those galaxies. Data to higher SFRs in both age ranges show good agreement with the SLUG simulations. 

Due to the short time period of the $<$10~Myr age range there should be little effect from cluster destruction, but there are five upper limit outliers that fall significantly below the expected relation in Figure 7. However, these galaxies (UGC8508, NGC4163, NGC3741, UGC5139, and UGC8201) all fall within the expected scatter from the stochastic simulations. The agreement of these outliers with stochastic simulations supports the "size-of-sample" explanation of \cite{larsen02} and \cite{Bastian08}'s conclusion that no outlier should significantly fall below the trend.

There are two galaxies that fall significantly above the $M_V^{\rm brightest}-SFR$ SLUG simulations: UGC5336 in the $<$10~Myr age range and UGC5692 in the $<$100~Myr age range. However, UGC5692 is shown not to be an outlier if the $<$10~Myr age range data is used as discussed in Section 4.3. The $<$10~Myr outlier, UGC5336, shows moderate agreement with the $M_V^{\rm brightest}-SFR$ trend, but shows poor agreement with the simulations. This cluster is a 98.2\% outlier, i.e. 982 of 1000 SLUG simulations produced clusters fainter than the observed cluster. 

\begin{figure}[h!]
  \begin{center}
  \includegraphics[scale=0.5]{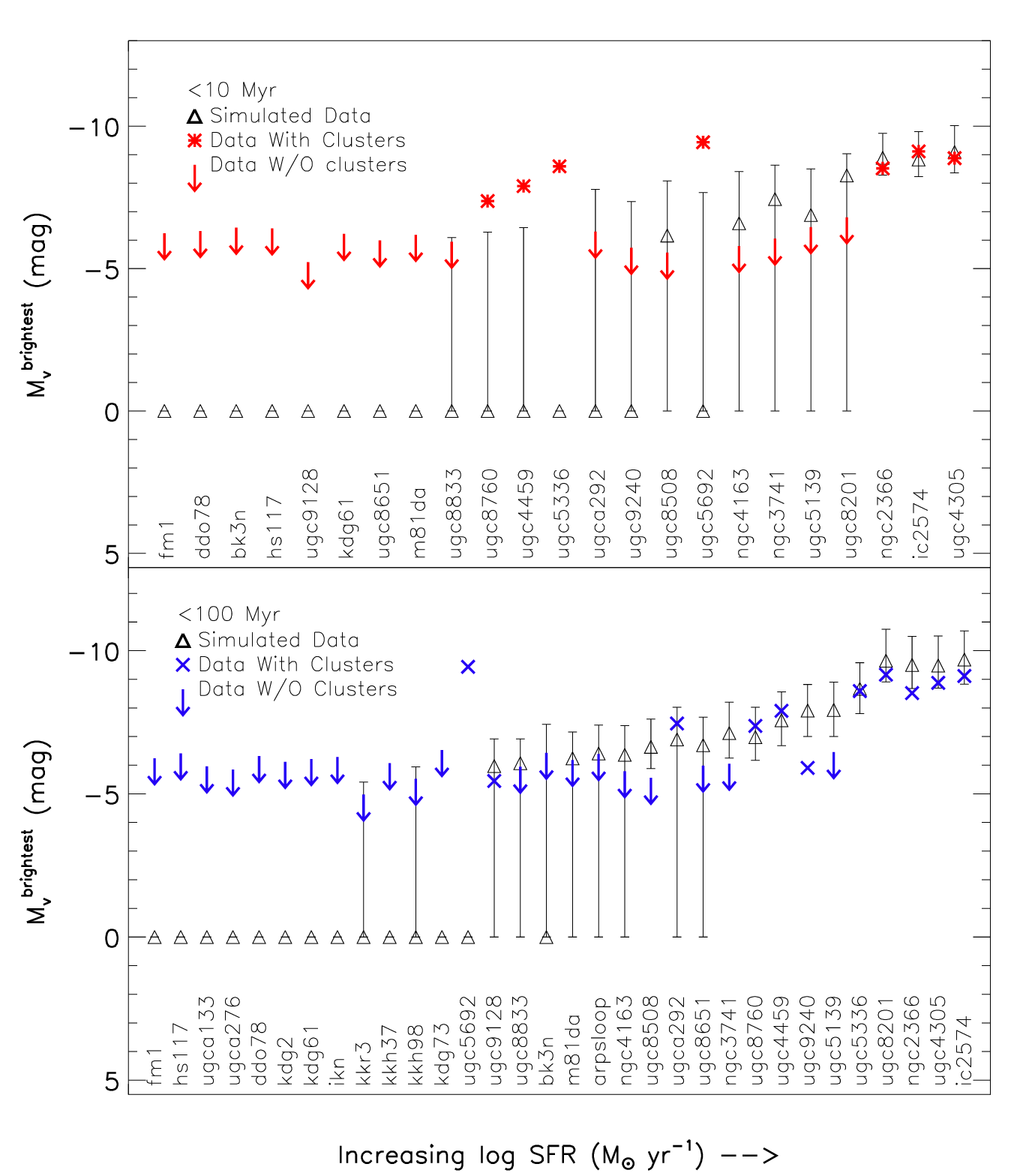}
  \caption{The simulated brightest cluster ($M_V^{\rm brightest}$) results from SLUG, where the top and bottom panels are the $<$10~Myr and $<$100~Myr age ranges, respectively. Each galaxy is presented on the $x$-axis evenly spaced and sorted by SFR, with increasing values to the right. The open triangles are the SLUG simulated galaxies and the error bars are the 68.3\% confidence inertvals of the 1000 realizations of the simulations. The blue X's are the observed data and the downward arrows are the absolute $V$ magnitude of the completeness limit for galaxies  with no detectable clusters. The majority of the observed data from both age ranges fall within the expected scatter from the SLUG simulations, and therefore show good agreement. However, there are a few significant outliers above and below the simulations which might indicate a true variation in the cluster formation of these galalxies.}
  \label{fig:figure}
   \end{center}
\end{figure}  


\subsection{Effects of Incompleteness}
Our completeness limit is only an approximate one and therefore it is possible that this limit can have an effect on the conclusions supported by our analysis. However, we will show in this section that the basic results are independent of the chosen completeness limit. 

The aspect of our analysis most affected by the completeness limit is the upper limit data points. The upper limits represent galaxies without cluster detections and are computed directly from the completeness limit. The $M_V^{\rm brightest}$ upper limits are computed as the completeness limit at the distance of the galaxies without cluster detections and the $\Gamma$ upper limits are computed as the total mass of clusters below the completeness limit. 

The two main conclusions of our analysis are 1) true variations in the cluster formation of dwarf galaxies and 2) that a mass-dependent cluster destruction model is dominant in our sample. The variation of cluster formation in dwarf galaxies is supported by the significant outliers in the $\Gamma$-$\Sigma_{SFR}$ relation not accounted for by stochasticity. The supporting data are comprised of galaxies \emph{with} cluster detections  (e.g., UGC5336, UGC4459 and NGC2366) and no upper limits. The cluster destruction conclusions, on the other hand, are supported by some upper limits. However, only four of the eleven galaxies used to draw these conclusions are upper limit data points. If the upper limits are ignored there would still be seven galaxies with cluster detections showing significantly low total cluster mass all of which show evidence for the most massive clusters surviving.

\section{Summary}
We have quantified the star formation and cluster formation properties of a large sample of dwarf galaxies. We use UBVRI data to derive the ages and masses of the cluster sample. Due to the low mass nature of the clusters in our sample we took extra care in the cluster parameter fitting using a Bayesian technique which specifically accounts for the stochastic sampling of stellar masses in low-mass clusters. 

The cluster sample was compared to previous studies by examining the $\Gamma$-$\Sigma_{SFR}$ and $M_V^{\rm brightest}$-SFR relationships in the age ranges $<$10~Myr and $<$100~Myr. This work provides the first examination of the these relationships in the low SFR regime. We found considerable scatter in our data compared to the $\Gamma$-$\Sigma_{SFR}$ relationship proposed by \cite{goddard10}. We also found moderate scatter in the $M_V^{\rm brightest}$-SFR relation for our dwarf galaxies. In the $M_V^{\rm brightest}$-SFR relationship some of our galaxies that lack clusters (upper limits) have predicted cluster magnitudes 3--4 magnitudes brighter than our upper limits. In the $\Gamma$-$\Sigma_{SFR}$ relationship NGC2366 shows a significant deviation to lower $\Gamma$ values and has also been found to have unusually low cluster formation by \cite{billett02}. Also, high $\Gamma$ deviations are seen in UGC5336, UGC4459, UGC8760, and UGC5692.

We model the expected scatter resulting from stochastically populating the CMF of our cluster sample using the SLUG software. We find that some of the outliers from the $\Gamma$-$\Sigma_{SFR}$ and $M_V^{\rm brightest}$-SFR relations are accounted for by the stochastic scatter, but there are galaxies which show significant deviations beyond the stochastic scatter. The large outliers in the $M_V^{\rm brightest}$-SFR relationship that fall 3--4 magnitudes below their predicted brightest clusters are within the stochastic scatter. Also, both UGC5692 and UGC8760 show significant scatter from the $\Gamma$-$\Sigma_{SFR}$ relationship, but fall within the scatter seen in the SLUG simulations. NGC2366, on the other hand, shows significant deviations from the $\Gamma$-$\Sigma_{SFR}$ relationship and the SLUG simulations in the $<$100~Myr age range. Furthermore, SLUG produced zero simulated cluster populations with $\Gamma$ values as high as the observed data points for UGC5336 and UGC4459 in the $<$10~Myr age range. Inspection of the resolved star CMD of each galaxy's clusters and H$\alpha$ images show correct age estimates. The cluster properties of these galaxies suggest that true variations in the clustered star formation are taking place in these low star formation rate galaxies.

Although we find that the majority of individual galaxies in the $\Gamma$-$\Sigma_{SFR}$ relationship show moderate agreement ($\sim2\sigma$)with the stochastic modeling of SLUG in the $<$100~Myr age range, the galaxy sample as a whole shows systematic $\Gamma$ values lower than the SLUG simulations. These lower values can be attributed to cluster destruction effects, since this age range will be affected more by cluster destruction. We constrain cluster destruction models by testing the mass-dependent model prediction that massive clusters preferentially survive cluster destruction. If a mass-dependent model is dominant in our sample, the galaxies with low $\Gamma$ values (i.e. low total cluster mass) should contain the brightest (i.e. most massive) cluster predicted by the $M_V^{\rm brightest}$-SFR relationship.

In our dwarf galaxy sample, 11 galaxies with significant scatter to lower $\Gamma$ values in the $<$100~Myr age range all were found to have good to moderate agreement with the brightest cluster predicted by the $M_V^{\rm brightest}$-SFR relationship. In other words, galaxies exhibiting a dearth of total cluster mass in an age range where cluster destruction becomes important all show evidence that the most massive clusters preferentially survive. This suggests that a mass-dependent cluster destruction model is dominant in our sample of dwarf galaxies.

We thank our referee for the insightful and constructive comments. Support for this work was provided by in part by NASA through grant number GO-10915 under the contract NASA 5-26555 and contract 1336000, and the Wyoming NASA Space Grant Consortium Grant \#NNX10A095H. This work is based on observations made with the NASA/ESA Hubble Space Telescope, obtained from the data archive at the Space Telescope Science Institute. This research has made use of the NASA/IPAC Infrared Science Archive and the NASA/IPAC Extragalactic Database (NED), which are both operated by the Jet Propulsion Laboratory, California Institute of Technology, under contract with the National Aeronautics and Space Administration. 

\bibliographystyle{apj}   
\bibliography{clusters}  

\newpage

\begin{appendix}
\end{appendix}

\begin{center}
\begin{deluxetable}{ccccccccccc}
\tabletypesize{\footnotesize}
\tablecolumns{11} 
\tablewidth{0pt} 
\tablecaption{Cluster Parameters}
\tablehead{\colhead{Galaxy}      		& 
	   \colhead{Cluster}     		& 
	   \colhead{RA}          		& 
	   \colhead{Dec}         		& 
	   \colhead{U}           		& 
	   \colhead{B}           		& 
	   \colhead{V}           		& 
	   \colhead{R}           		& 
	   \colhead{I}           		& 
	   \colhead{log(Age)}    		&
	   \colhead{log(Mass)}   		\\
                       		&
	   \colhead{(\#)}        		&
           \colhead{(J2000.0)}   		&        
           \colhead{(J2000.0)}   		&        
	   \colhead{(mag)}       		&
	   \colhead{(mag)}       		&
	   \colhead{(mag)}       		&
	   \colhead{(mag)}       		&
	   \colhead{(mag)}       		&
	   \colhead{(yr)}       		&
	   \colhead{($M_{\odot}$)}	        }
\startdata
   ngc2366 &    03 & 07~28~40.4 & 69~10~09.2 & 21.01$\pm$0.04& 21.45$\pm$0.06& 21.01$\pm$0.02& 20.82$\pm$0.06& 20.28$\pm$0.03 & 7.6 & 3.4 \\  
   ngc2366 &    05 & 07~29~00.8 & 69~14~08.2 & 21.15$\pm$0.09& 21.78$\pm$0.10& 21.85$\pm$0.13& 21.82$\pm$0.26& 21.17$\pm$0.14 & 7.6 & 3.4 \\  
   ngc2366 &    06 & 07~29~15.3 & 69~14~36.2 & 20.44$\pm$0.03& 20.40$\pm$0.05& 19.97$\pm$0.01& 19.37$\pm$0.12& 18.71$\pm$0.01 & 7.4 & 3.7 \\  
   ngc2366 &    18 & 07~29~04.0 & 69~13~01.3 & 17.92$\pm$0.03& 19.13$\pm$0.05& 19.01$\pm$0.02& 19.11$\pm$0.12& 19.66$\pm$0.03 & 6.5 & 3.4 \\  
   ugc4305 &    03 & 08~18~54.5 & 70~42~38.3 & 20.17$\pm$0.11& 20.71$\pm$0.08& 20.63$\pm$0.06& 20.49$\pm$0.09& 20.18$\pm$0.06 & 7.6 & 3.8 \\  
   ugc4305 &    23 & 08~18~53.1 & 70~41~59.7 & 20.63$\pm$0.14& 20.76$\pm$0.08& 20.68$\pm$0.08& 19.82$\pm$0.06& 19.27$\pm$0.04 & 7.4 & 3.7 \\  
   ugc4305 &    37 & 08~18~54.9 & 70~43~13.0 & 19.20$\pm$0.05& 20.37$\pm$0.06& 20.31$\pm$0.03& 20.29$\pm$0.07& 20.44$\pm$0.08 & 6.8 & 3.7 \\  
   ugc4305 &    11 & 08~19~26.6 & 70~41~57.1 & 17.95$\pm$0.04& 18.94$\pm$0.05& 18.77$\pm$0.01& 18.87$\pm$0.04& 19.15$\pm$0.01 & 6.8 & 3.7 \\  
   ugc4305 &    21 & 08~19~28.8 & 70~43~05.3 & 18.38$\pm$0.04& 19.61$\pm$0.05& 19.37$\pm$0.01& 19.20$\pm$0.04& 19.69$\pm$0.03 & 6.5 & 3.9 \\  
   ugc4305 &    22 & 08~19~28.5 & 70~42~46.4 & 18.09$\pm$0.04& 19.07$\pm$0.06& 19.24$\pm$0.04& 18.92$\pm$0.05& 19.09$\pm$0.02 & 7.2 & 4.4 \\  
   ugc4459 &    11 & 08~34~07.8 & 66~10~51.0 & 19.11$\pm$0.07& 20.23$\pm$0.04& 19.89$\pm$0.03& 19.90$\pm$0.06& 20.32$\pm$0.05 & 6.3 & 3.5 \\  
   ugc5336 &    02 & 09~57~39.8 & 69~03~23.7 & 19.06$\pm$0.12& 19.45$\pm$0.04& 19.20$\pm$0.01& 18.83$\pm$0.03& 18.72$\pm$0.01 & 7.0 & 3.7 \\  
   ugc5336 &    17 & 09~57~37.5 & 69~02~27.9 & 20.18$\pm$0.03& 20.65$\pm$0.02& 20.28$\pm$0.02& 20.11$\pm$0.05& 19.73$\pm$0.02 & 7.2 & 3.7 \\  
   ugc5336 &    28 & 09~57~28.4 & 69~02~44.0 & 20.60$\pm$0.04& 21.34$\pm$0.05& 20.67$\pm$0.04& 20.55$\pm$0.07& 20.26$\pm$0.05 & 7.4 & 3.4 \\  
   ugc5336 &    34 & 09~57~33.1 & 69~03~20.3 & 21.12$\pm$0.07& 21.87$\pm$0.10& 21.62$\pm$0.10& 22.07$\pm$0.28& 21.23$\pm$0.13 & 7.6 & 3.5 \\  
   ugc5336 &    35 & 09~57~31.1 & 69~03~24.1 & 21.88$\pm$0.14& 22.94$\pm$0.10& 22.69$\pm$0.10& 21.36$\pm$0.10& 22.31$\pm$0.10 & 7.6 & 3.1 \\  
   ugc5336 &    37 & 09~57~26.2 & 69~03~08.9 & 21.26$\pm$0.13& 22.47$\pm$0.08& 22.29$\pm$0.08& 21.62$\pm$0.16& 22.24$\pm$0.14 & 6.1 & 3.1 \\  
    ic2574 &    02 & 10~28~29.1 & 68~24~05.5 & 21.28$\pm$0.11& 21.89$\pm$0.14& 22.16$\pm$0.07& 21.29$\pm$0.10& 21.52$\pm$0.09 & 6.1 & 3.1 \\  
    ic2574 &    10 & 10~28~21.0 & 68~24~30.8 & 20.43$\pm$0.07& 20.66$\pm$0.09& 20.64$\pm$0.04& 20.15$\pm$0.06& 19.92$\pm$0.04 & 7.6 & 3.8 \\  
    ic2574 &    11 & 10~28~17.5 & 68~24~28.7 & 21.71$\pm$0.14& 22.13$\pm$0.13& 21.65$\pm$0.06& 21.15$\pm$0.07& 20.72$\pm$0.06 & 7.4 & 3.2 \\  
    ic2574 &    13 & 10~28~14.2 & 68~24~40.3 & 21.53$\pm$0.10& 22.09$\pm$0.15& 22.09$\pm$0.10& 21.91$\pm$0.16& 21.58$\pm$0.13 & 7.8 & 3.5 \\  
    ic2574 &    33 & 10~28~20.6 & 68~26~19.1 & 20.19$\pm$0.02& 20.55$\pm$0.02& 20.27$\pm$0.02& 20.26$\pm$0.03& 19.81$\pm$0.02 & 7.2 & 3.7 \\  
    ic2574 &    02 & 10~27~47.6 & 68~21~38.6 & 20.83$\pm$0.04& 20.88$\pm$0.05& 20.65$\pm$0.01& 20.24$\pm$0.03& 20.03$\pm$0.01 & 7.4 & 3.2 \\  
    ic2574 &    04 & 10~28~06.1 & 68~22~16.1 & 21.31$\pm$0.06& 21.60$\pm$0.05& 21.77$\pm$0.05& 21.55$\pm$0.06& 21.41$\pm$0.07 & 7.8 & 3.7 \\  
    ic2574 &    08 & 10~27~41.1 & 68~22~04.7 & 21.99$\pm$0.16& 21.83$\pm$0.09& 22.95$\pm$0.11& 21.89$\pm$0.15& 22.48$\pm$0.14 & 7.8 & 3.4 \\  
    ic2574 &    12 & 10~27~55.0 & 68~23~25.0 & 21.61$\pm$0.07& 21.98$\pm$0.05& 21.19$\pm$0.02& 20.81$\pm$0.04& 20.17$\pm$0.02 & 7.4 & 3.2 \\  
    ic2574 &    01 & 10~28~47.9 & 68~25~30.2 & 20.67$\pm$0.09& 21.16$\pm$0.10& 21.30$\pm$0.04& 21.09$\pm$0.08& 20.12$\pm$0.02 & 7.6 & 3.7 \\  
    ic2574 &    12 & 10~28~42.1 & 68~26~41.7 & 20.31$\pm$0.05& 21.02$\pm$0.07& 21.38$\pm$0.03& 21.15$\pm$0.07& 20.98$\pm$0.06 & 6.8 & 3.4 \\  
    ic2574 &    25 & 10~28~45.3 & 68~28~02.1 & 19.35$\pm$0.09& 19.91$\pm$0.20& 19.86$\pm$0.12& 19.85$\pm$0.25& 19.54$\pm$0.21 & 7.6 & 4.2 \\  
    ic2574 &    72 & 10~28~44.6 & 68~28~11.4 & 17.87$\pm$0.05& 18.53$\pm$0.07& 18.85$\pm$0.03& 18.41$\pm$0.06& 18.16$\pm$0.05 & 6.5 & 3.9 \\  
    ic2574 &    73 & 10~28~44.6 & 68~28~07.8 & 17.71$\pm$0.04& 18.33$\pm$0.06& 18.79$\pm$0.03& 18.55$\pm$0.05& 18.28$\pm$0.04 & 6.8 & 4.1 \\  
    ic2574 &    79 & 10~28~37.3 & 68~27~57.2 & 18.69$\pm$0.04& 19.55$\pm$0.06& 19.50$\pm$0.02& 19.59$\pm$0.04& 19.83$\pm$0.04 & 7.2 & 4.2 \\  
   ugc5692 &    13 & 10~30~35.0 & 70~37~08.3 & 19.93$\pm$0.11& 18.89$\pm$0.03& 18.46$\pm$0.03& 18.93$\pm$0.05& 18.64$\pm$0.04 & 6.8 & 3.1 \\  
   ugca292 &    10 & 12~38~40.0 & 32~46~02.4 & 19.65$\pm$0.04& 20.49$\pm$0.04& 20.34$\pm$0.04& 20.19$\pm$0.05& 20.39$\pm$0.03 & 7.6 & 3.9 \\  
   ugc8201 &    08 & 13~~6~30.3 & 67~41~51.1 & 20.26$\pm$0.04& 20.60$\pm$0.04& 20.51$\pm$0.03& 20.30$\pm$0.05& 20.03$\pm$0.03 & 7.8 & 4.2 \\  
   ugc8201 &    09 & 13~~6~22.6 & 67~42~01.2 & 19.49$\pm$0.03& 19.92$\pm$0.03& 19.90$\pm$0.02& 19.81$\pm$0.04& 19.52$\pm$0.03 & 7.6 & 4.4 \\  
   ugc8201 &    10 & 13~~6~18.0 & 67~42~12.8 & 18.62$\pm$0.02& 19.08$\pm$0.02& 19.14$\pm$0.02& 18.79$\pm$0.03& 18.76$\pm$0.01 & 7.4 & 4.5 \\  
   ugc8201 &    11 & 13~~6~17.6 & 67~41~55.1 & 21.94$\pm$0.12& 22.85$\pm$0.18& 22.50$\pm$0.21& 22.81$\pm$0.79& 22.09$\pm$0.30 & 7.6 & 3.4 \\  
   ugc8201 &    12 & 13~~6~16.5 & 67~42~04.8 & 20.69$\pm$0.09& 21.04$\pm$0.08& 21.06$\pm$0.06& 20.00$\pm$0.05& 20.60$\pm$0.07 & 7.2 & 3.7 \\  
   ugc8201 &    34 & 13~~6~29.9 & 67~42~13.9 & 20.73$\pm$0.05& 21.23$\pm$0.05& 21.22$\pm$0.05& 20.91$\pm$0.07& 20.81$\pm$0.06 & 7.6 & 3.9 \\  
   ugc8201 &    38 & 13~~6~25.9 & 67~42~03.7 & 21.30$\pm$0.09& 21.09$\pm$0.06& 21.77$\pm$0.10& 21.07$\pm$0.10& 21.21$\pm$0.10 & 7.4 & 3.4 \\  
   ugc8760 &    16 & 13~50~51.7 & 38~01~18.6 & 19.15$\pm$0.02& 20.59$\pm$0.07& 20.17$\pm$0.08& 20.75$\pm$0.11& 20.66$\pm$0.16 & 6.5 & 3.1 \\  
   ugc9128 &    09 & 14~15~55.0 & 23~03~02.7 & 20.44$\pm$0.06& 21.27$\pm$0.08& 21.27$\pm$0.12& 21.71$\pm$0.19& 21.58$\pm$0.24 & 7.2 & 3.1 \\  
   ugc9240 &    12 & 14~24~42.9 & 44~31~42.4 & 20.22$\pm$0.09& 21.26$\pm$0.19& 21.32$\pm$0.19& 20.65$\pm$0.24& 20.93$\pm$0.22 & 7.6 & 3.7 \\  
\enddata
\tablecomments{The photometric properties of the cluster sample with an age less than 100~Myr.We report each cluster's identification number, right ascension, declination, UBVRI magnitude with error, the fitted age, and the fitted mass.}
\end{deluxetable}
\end{center}

\end{document}